\documentclass[a4paper,11pt]{article}
\usepackage{jheppub}
\usepackage[T1]{fontenc}

\usepackage{comment}
\usepackage{multirow}
\usepackage{dcolumn}
\usepackage{bm}
\usepackage{setspace}
\usepackage{graphicx}
\usepackage{xspace}

\usepackage{dcolumn}
\usepackage{bm}

\newenvironment{cfigure1c}[1][tbp]{\begin{figure*}[#1]\centering}{\end{figure*}}

\newcommand{\ppbar}  {\ensuremath{p\bar{p}}\xspace}
\newcommand{\ttbar}  {\ensuremath{t\bar{t}}\xspace}

\newcommand{\et}     {\ensuremath{E_{T}}\xspace}

\newcommand{\pt}     {\ensuremath{p_{T}}\xspace}

\newcommand{\met}    {\ensuremath{\text{\raisebox{.3ex}{$\not$}}\et}\xspace}

\newcommand{\chisq}  {\ensuremath{\chi^{2}}\xspace}

\newcommand{\mtop}    {\ensuremath{\mathrm{M}_{\text{top}}}\xspace}

\newcommand{\mtreco}     {\ensuremath{m_{t}^{\text{reco}}}\xspace}
\newcommand{\afb}     {\ensuremath{A_{\text{FB}}}\xspace}

\newcommand{\genunit}[2]{\ensuremath{#1~\mathrm{#2}}\xspace}

\newcommand{\gev}[1]    {\genunit{#1}{GeV}}

\newcommand{\gevc}[1]   {\ensuremath{#1~\mathrm{GeV}/c}}
\newcommand{\tevcc}[1]  {\ensuremath{#1~\mathrm{TeV}/c^{2}}}
\newcommand{\gevcc}[1]  {\ensuremath{#1~\mathrm{GeV}/c^{2}}}

\newcommand{\invfb}[1]  {\ensuremath{#1~\mathrm{fb}^{-1}}}

\begin{document}

\title{\boldmath Improving Combinatorial Ambiguities of \ttbar Events Using Neural Networks}

\author[1]{Ji Hyun Shim and Hyun Su Lee \note{Corresponding Author}}
\affiliation{Department of Physics, Ewha Womans University, Seoul, 120-750, Korea}
\emailAdd{turquoise@ewhain.net}
\emailAdd{hyunsulee@ewha.ac.kr}

\abstract{
We present a method for resolving the combinatorial issues in the \ttbar lepton+jets events 
occurring at the Tevatron collider. 
By incorporating multiple information into an artificial neural network, we introduce a 
novel event reconstruction method for such events. 
We find that this method significantly reduces the number of combinatorial ambiguities. 
Compared to the classical reconstruction method, our method provides significantly higher  
purity with same efficiency. 
We illustrate the reconstructed observables for the realistic top-quark mass and the 
forward-backward asymmetry measurements. 
A Monte Carlo study shows that our method provides meaningful improvements in the 
top-quark measurements using same amount of data as other methods. 
}

                              
\maketitle
\flushbottom 

\section{Introduction}
The top quark was discovered in 1995 at the Fermilab Tevatron \ppbar 
Collider~\cite{top_discovery}. This recently discovered quark is the heaviest known elementary 
particle~\cite{pdg}. Its large mass may strongly couple with the electroweak symmetry 
breaking~\cite{sbhigg,sbreview}, and therefore, the top quark is usually treated differently 
from the other light quarks in many new physics models. This suggests that many searches focus 
on the top quark signature~\cite{tsearch}. Recent observations of the charge forward-backward 
asymmetry~(\afb) at the Tevatron collider may provide evidence for the involvement of this 
new physical signature in \ttbar production~\cite{cdf_afb,d0_afb}. 
However, because of the limited number of \ttbar events at Tevatron, updated measurements 
using the full Tevatron data at the Collider Detector at Fermilab (CDF) still do not 
confirm whether they have identified a new physical signature or if the result is a 
statistical fluctuation of the standard model (SM) process~\cite{cdf_afbfull}.
Because the Large Hadron Collider (LHC) is a $pp$ collider, it is difficult to probe all of
the possible scenarios of new physics involved in the top-quark \afb. 
It is, therefore, necessary to use the currently available Tevatron data as efficiently 
as possible. 

Similar issues have occurred with the top-quark mass~(\mtop) measurements. The top-quark mass 
is a fundamental parameter of the SM, and it is tightly related to the $W$-boson mass and 
Higgs-boson mass via electroweak radiative corrections~\cite{elfit,gfit}. Therefore, precision 
measurements of \mtop and the $W$-boson indirectly provide important constraints on the Higgs 
boson mass. Compared with the directly-observed Higgs-boson mass~\cite{higgs}, precision 
measurement of \mtop can be important for understanding the SM. 
Even though LHC experiments obtained \ttbar events that were more than two orders of magnitude 
larger than the Tevatron events, the Tevatron measurements give the most precise \mtop results 
to date~\cite{tevave,lhcave}. Because of the well-known detectors as well as the much 
smaller instantaneous luminosity of the collisions, the Tevatron measurements usually 
have smaller systematic uncertainties. Consequently, the dominant uncertainty of the Tevatron 
\mtop measurement is statistical uncertainty~\cite{tevave}, while the LHC measurement is 
dominated by systematic uncertainty~\cite{lhcave}. Therefore, efficiently utilizing the 
currently available Tevatron data is important for improving the precision of Tevatron 
measurements as well as world average of \mtop.  

In the SM, the top quark decays almost exclusively into a $W$ boson and a $b$ quark~\cite{pdg}. 
In \ttbar events, the lepton+jets decay channel is defined by the case where one $W$ boson 
decays leptonically into an electron or a muon plus a neutrino and the other $W$ decays 
hadronically into a pair of jets. Thus, the events in this channel contain one charged 
lepton, two $b$ quark jets, two light flavor quark jets, and one undetected neutrino. 
Many precise top-quark measurements including \mtop and \afb have been performed using 
lepton+jets events.
 
In the lepton+jets event, we have measured four final jets in the detector that should 
be matched with initial parton-level quarks. During event reconstruction, we should resolve 
these combinatorial ambiguities. The rate of correctly matching events for all four jets in 
the kinematic reconstruction method, which is commonly used in hadron collider 
experiments~\cite{cdf_oldtop,d0_fitter,atlas_mass,cms_mass}, was only 18--47\% depending 
on the $b$-tagging category after poorly reconstructed events were rejected by requiring 
that the minimum $\chi^2$ be less than nine~\cite{cdf_oldtop}. Other events use imperfect  
\ttbar reconstruction from incorrect matching between jets and quarks. As shown in 
Ref.~\cite{cdf_oldtop}, incorrectly matching events resulted much poorer resolution in 
the distribution of observables such as the reconstructed top-quark mass and $W$-boson mass. 
If we develop an improved method for resolving the combinatorial ambiguities, the majority
of the top-quark measurements can be significantly improved. In this paper, we report a 
novel technique for improving the combinatorial ambiguities of \ttbar lepton+jets events 
using a multivariable neural network~(NN) technique. This method is directly applicable 
to experimental measurements of the top-quark such as \mtop and \afb. 

There are a few studies that discuss the combinatoric ambiguities at hadron 
colliders~\cite{comb1,comb2,comb3}. However, these studies focus on new physics processes in pair 
production, and both particle and anti-particle decay into invisible particles that could be 
dark matter candidates. Therefore, the final state contains two invisible particles together 
with the visible particles. These methods can be directly applicable for the dilepton decay 
channel of \ttbar events due to the two invisible neutrinos, but they are not feasible for 
lepton+jets decay events. Furthermore, those studies consider model-independent analysis 
because we may not exactly know what is underlying the new physics seen in the data. 
On the other hand, \ttbar productions and decays are very precisely tested in the SM 
framework~\cite{pdg}.  We can therefore use model-dependent analysis to maximize the 
information for resolving the combinatoric ambiguities. One very useful technique for incorporating 
multiple information is an artificial NN method~\cite{NNhep1,NNhep2}. 
Here, we propose the use of the artificial NN for resolving the combinatoric ambiguities in
\ttbar events. 

\section{\ttbar events and its reconstruction} 
For the study of \ttbar event reconstruction, we generated a simulated \ttbar signal sample 
with \mtop = \gevcc{173}~\cite{tevave,lhcave} using the leading order~(LO) Monte-Carlo~(MC) 
generator {\sc madgraph/madevent}~\cite{madgraph} package with {\sc pythia}~\cite{pythia} 
parton showering. The detector effects are produced with the fast detector simulation 
package~{\sc pgs}~\cite{pgs}. The detector resolution effects are simulated by the following 
parametrization:
$$\frac{\delta E}{E} = \frac{a}{\sqrt{E}}~ ~ \mathrm{for~ jets,}$$
$$\frac{\delta E}{E} = \frac{b}{\sqrt{E}} \oplus c~ ~ \mathrm{for~ leptons.}$$
As per the predefined values in the {\sc pgs} package, we let $a=0.8$, $b=0.2$, and $c=0.01$. 
The {\sc pgs} package can also quickly reconstruct each physical object such as leptons, jets, 
and missing transverse energy. In the simulation, the jets originating from $b$ quarks are 
tagged with approximately 40\% $b$-tagging efficiency.

To select the candidate events in the \ttbar lepton+jets channel, we require one charged 
lepton candidate with transverse momentum, $p_{T}$, greater than $\gevc{20}$. We also require 
missing transverse energy~(\met) to exceed \gev{20} and at least four jets with transverse 
energy, $E_{T}$, greater than $\gev{20}$. We further request that at least one jet is tagged 
as a $b$ quark.

We first attempt to reconstruct the \ttbar lepton+jets events using the kinematic reconstruction 
method applied in the CDF analyses~\cite{cdf_oldtop,cdf_fitter}. 
We build a $\chi^2$ formula to obtain the most probable combination that combines all 
measured quantities and known constraints. 
Our slightly method differs from the CDF description in that we directly use \met instead of 
the unclustered energy with approximately 40\% resolution because the unclustered energy is 
unavailable in the fast simulation. Therefore, we define $\chi^{2}$ for the kinematic fit 
as follows:
\begin{eqnarray}
\label{eq_chi2}
\chisq & = &
\Sigma_{i = \ell, 4 jets} {(p_T^{i,fit} - p_T^{i,meas})^2 / \sigma_i^2}  +  
\Sigma_{k = x,y} {(\nu_{T_k}^{fit} - {\mathrm{\raisebox{.3ex}{$\not$}}E}_{T_k}^{meas})^2 / \sigma_k^2} \nonumber \\
& + & {({M}_{jj} - {M}_{W})^2 / \Gamma_{W}^2}
 + {({M}_{\ell\nu} - {M}_{W})^2 / \Gamma_{W}^2} \nonumber \\
& + &  {\{{M}_{bjj} - {M}_{\text{top}}\}^2 / \Gamma_t^2} 
 +  {\{{M}_{b\ell\nu} - {M}_{\text{top}}\}^2 / \Gamma_t^2}\nonumber.
\end{eqnarray}
In this $\chi^2$ formula, the first term constrains the $\pt$ of the lepton and the four leading 
jets to their measured values within their respective uncertainties~(detector resolutions).
The second term constrains both transverse components of \met, $x$ and $y$, as well as those of 
the neutrino, $p_{x}$ and $p_{y}$. In the last four terms, the quantities 
${M}_{jj}, {M}_{\ell\nu}, {M}_{bjj}$, and ${M}_{b\ell\nu}$ refer to the invariant masses 
of the four-vector sum of the particles denoted in the subscripts. Here, ${M}_W$ and 
${M}_{\text{top}}$ are the masses of the $W$ boson~(\gevcc{80.4})~\cite{wmass} and
the top quark, respectively, and ${M}_{\text{top}}$ is determined during minimization of 
$\chi^2$. The total widths of the $W$ boson and the top quark are $\Gamma_W$ (\gevcc{2.1}) 
and $\Gamma_t$ (\gevcc{1.5}), respectively~\cite{pdg}.

Assuming that the four leading jets in the detector are products of the \ttbar decay, there are 
twelve possible jet-to-quark assignments. We perform a minimization for each assignment using 
a $\chi^2$ comparison. In the classical kinematic reconstruction method~($\chi^2$ method), 
the combination that has the lowest $\chi^2$ is selected as a candidate for correctly matched 
events. To understand the performance of the reconstruction methods, we study the true quark 
properties together with the reconstructed jet properties. If the distance, 
$\Delta R \equiv \sqrt{(\Delta\eta)^{2}+(\Delta\phi)^{2}}$, between a quark and a reconstructed 
jet is less than 0.4, the jet-to-parton assignment is correct. If all four quarks and jets 
have correct assignments, this event has correct matching. We then obtain a purity, which is 
a fraction of the correct matching events, of 35\% using the classical $\chi^2$ method for the 
SM \ttbar events. If we use the CDF requirement that $\chi^2$ must be less than 9, which 
has an event efficiency of 76\%, the purity is increased to 39\%. The values of efficiency and 
purity are consistent with those in the CDF results~\cite{cdf_oldtop}. 

Because there are 12 possible combinatorics, a purity about 35\% is acceptable. Moreover, under 
our assumption that the four leading jets are candidates for the four quarks, the maximum purity 
of the SM \ttbar sample is only 67\%. However, without losing efficiency, the purity 
can be improved by up to 90\%~($35\%\rightarrow67\%$). Even though $\chi^2$ is important for 
matching the jets and quarks, there are many additional variables that can be used to 
determine the correctly matching combinations. Therefore, we employ the multivariable NN 
for the event reconstruction of \ttbar lepton+jets events.
 
We use the results of the kinematic fitter for all possible combinatorics as the input for our 
neural network. Of the possible combinations, we choose a signal, which has all jets matched 
with the correct quarks, and background, where at least one jet is unmatched.  We then train 
the NN with various kinematic input variables that have discrimination powers between the 
signal and the background. To avoid any measurement bias and to maximize the discrimination
power, the choice of input variables depends on the measurements.  
The NN is constructed with the network implementation in the {\sc root} package~\cite{root}.

\section{\mtop measurement} 
\begin{cfigure1c}
\begin{tabular}{ccc}
\includegraphics[width=0.32\textwidth]{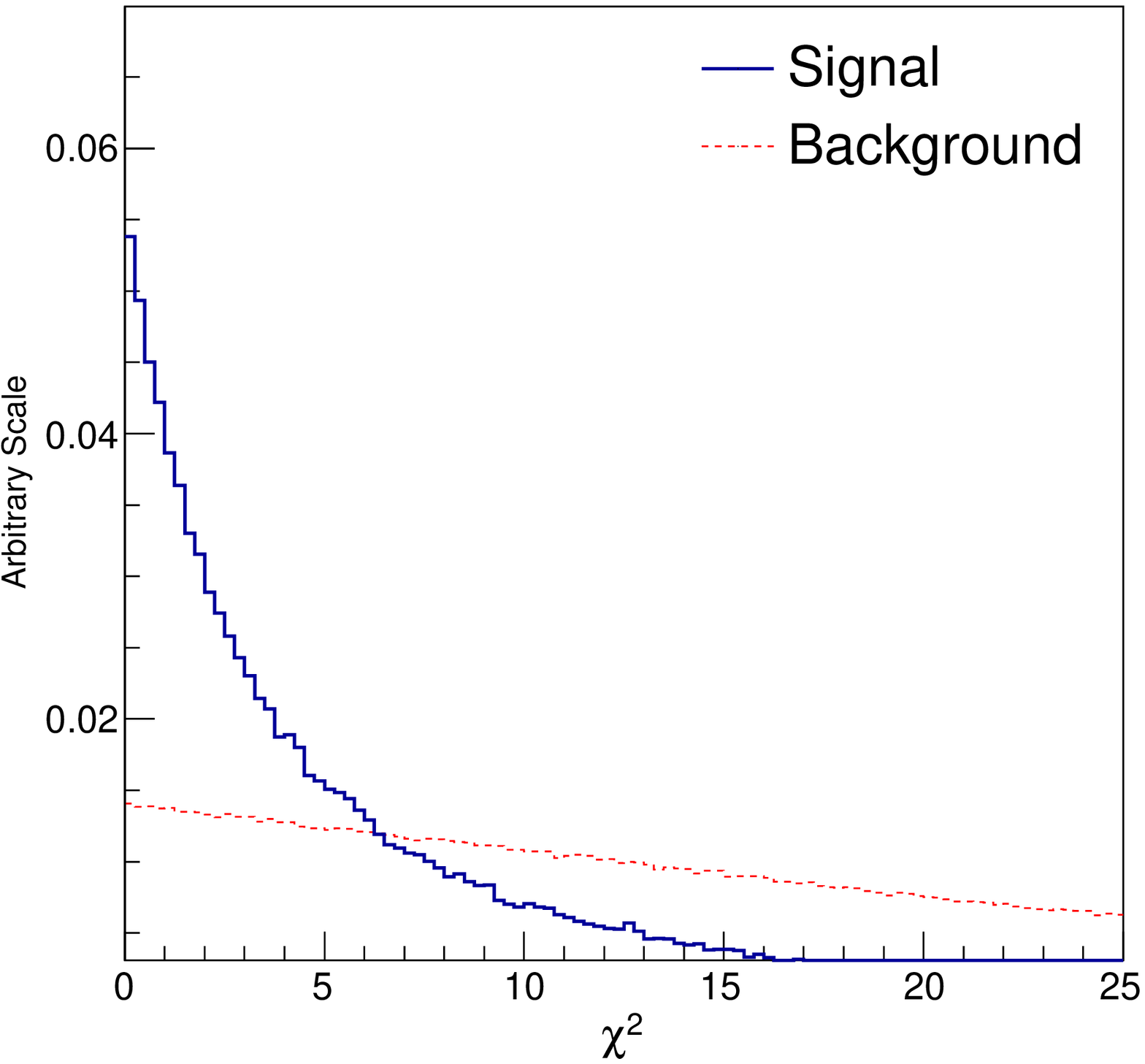}&
\includegraphics[width=0.32\textwidth]{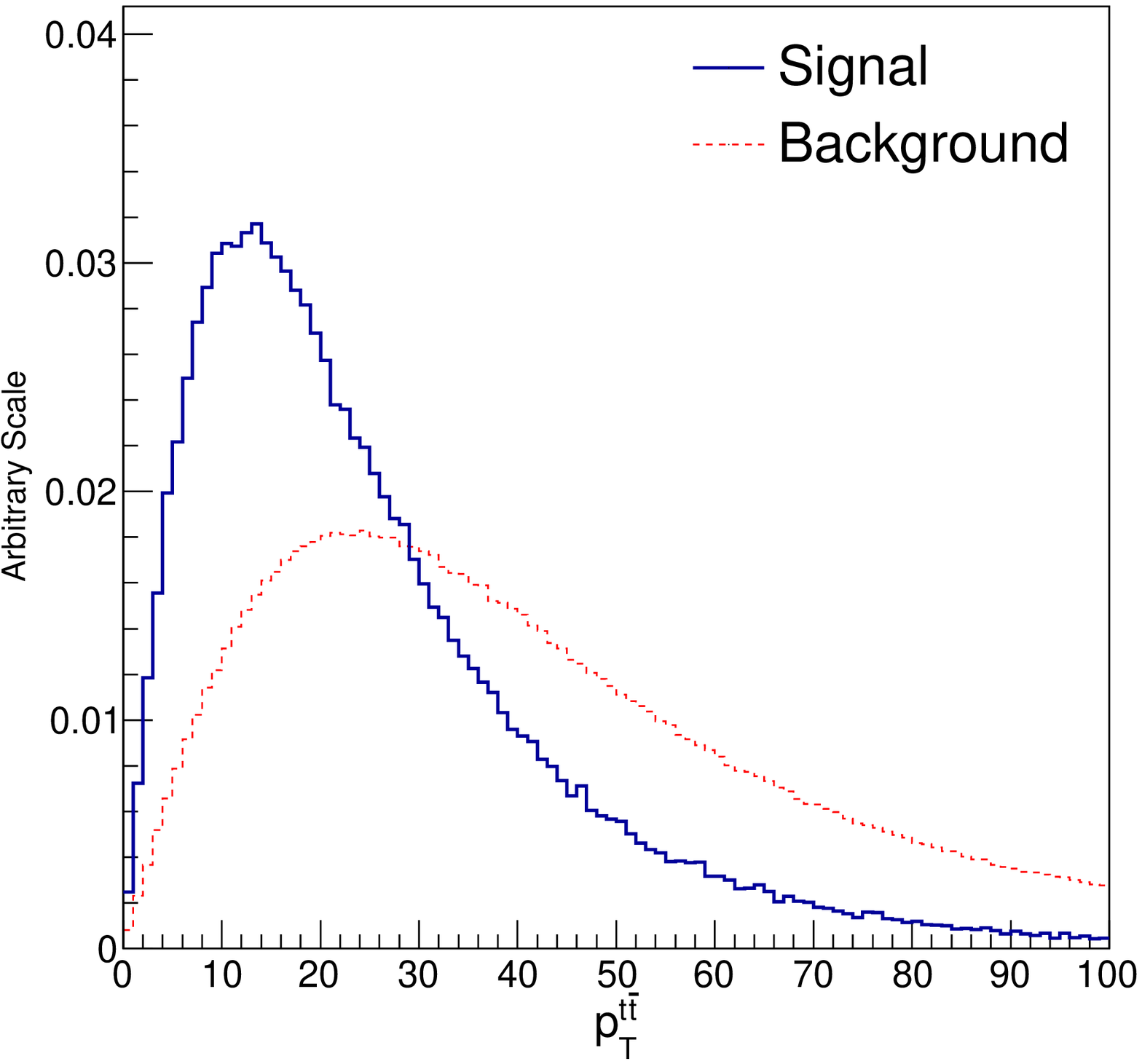}&
\includegraphics[width=0.32\textwidth]{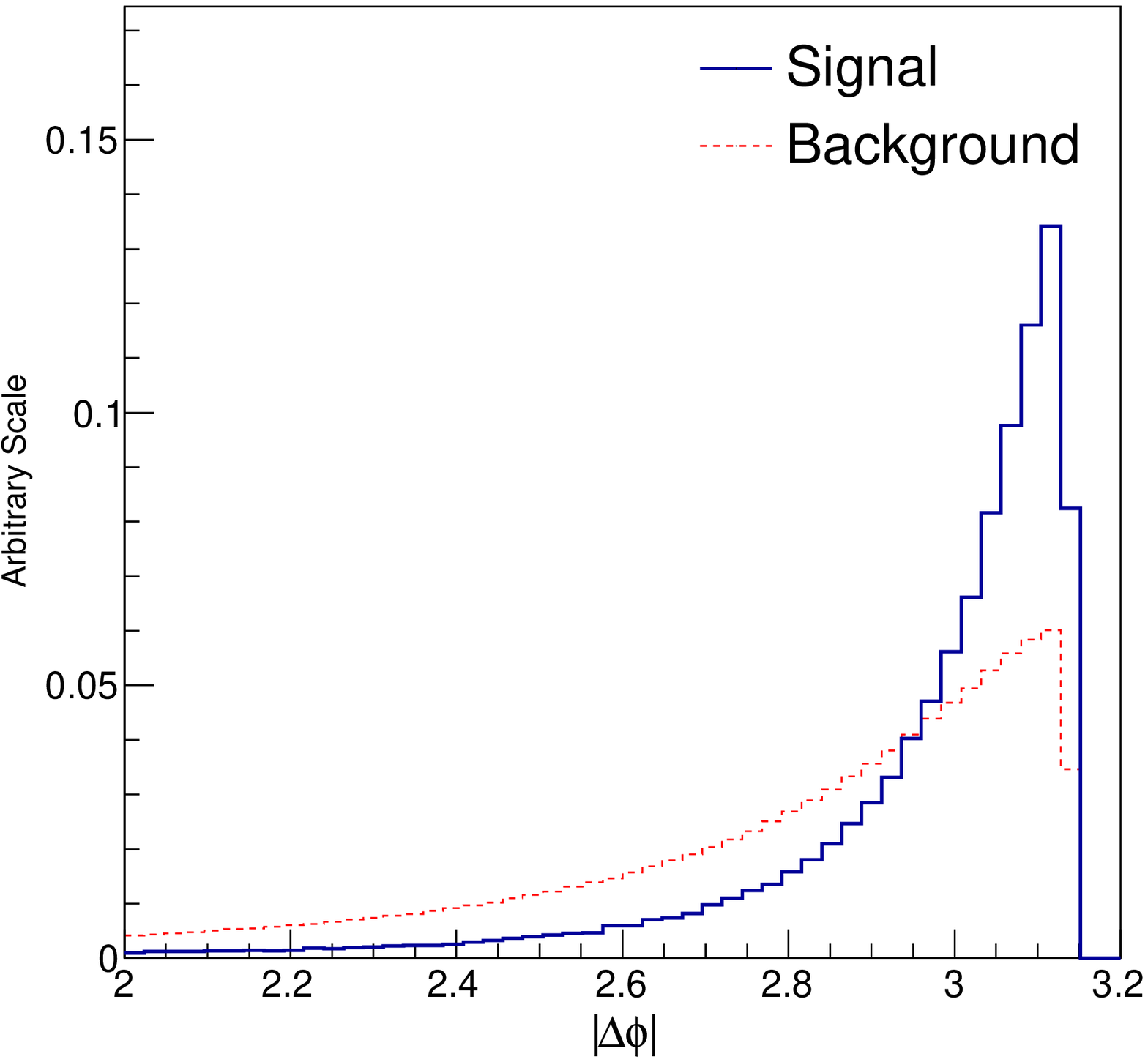}\\
\includegraphics[width=0.32\textwidth]{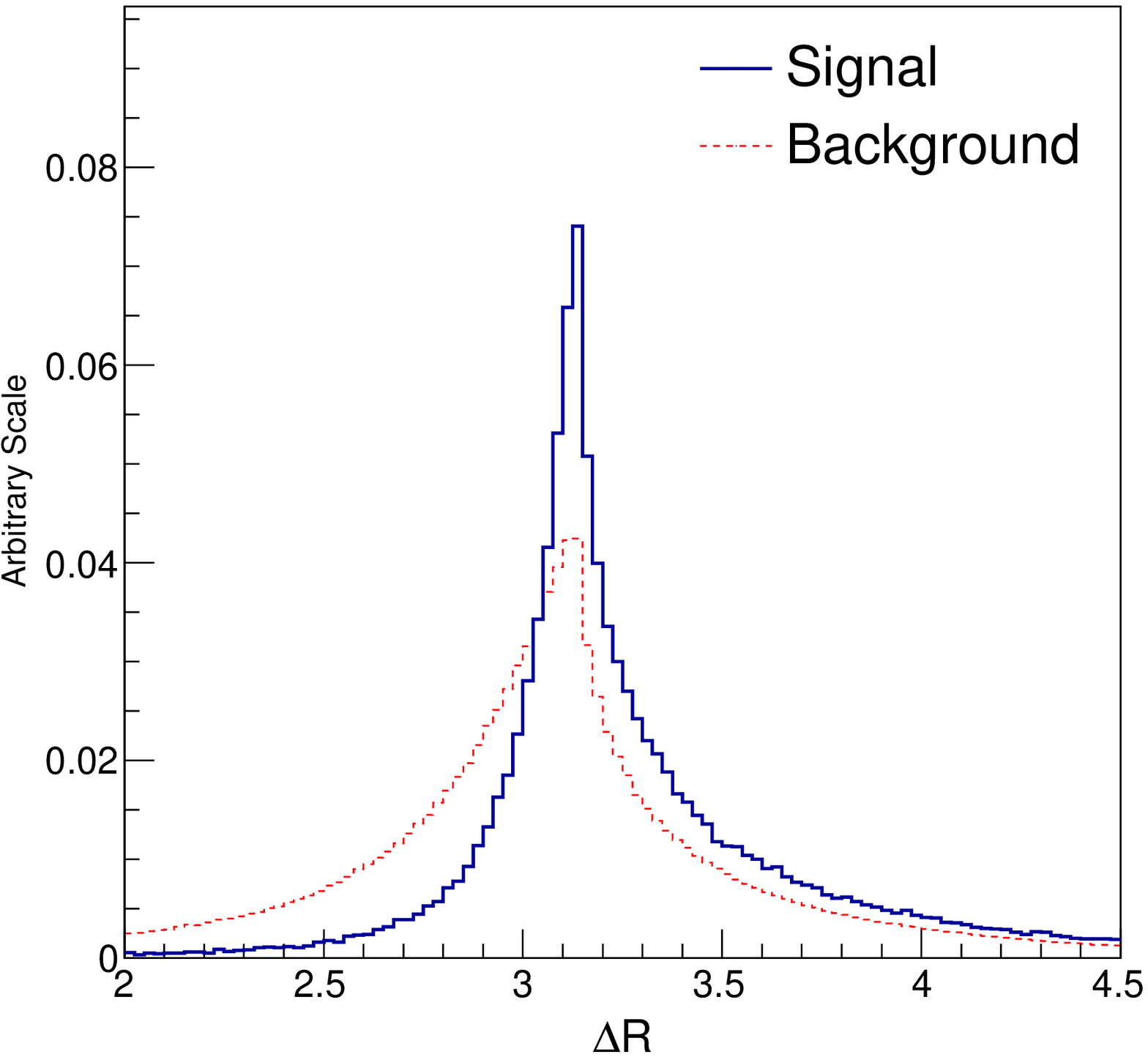}&
\includegraphics[width=0.32\textwidth]{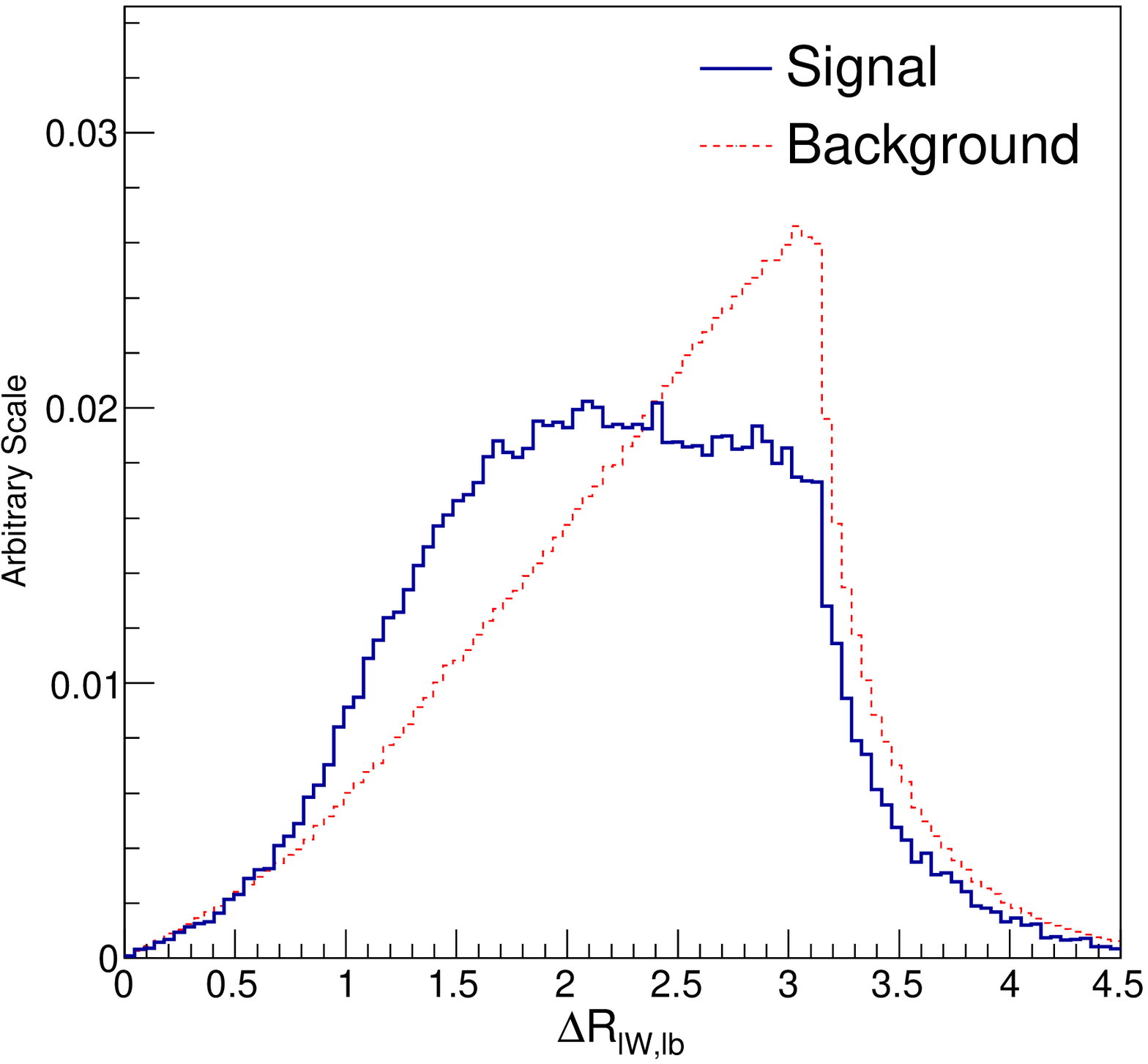}&
\includegraphics[width=0.32\textwidth]{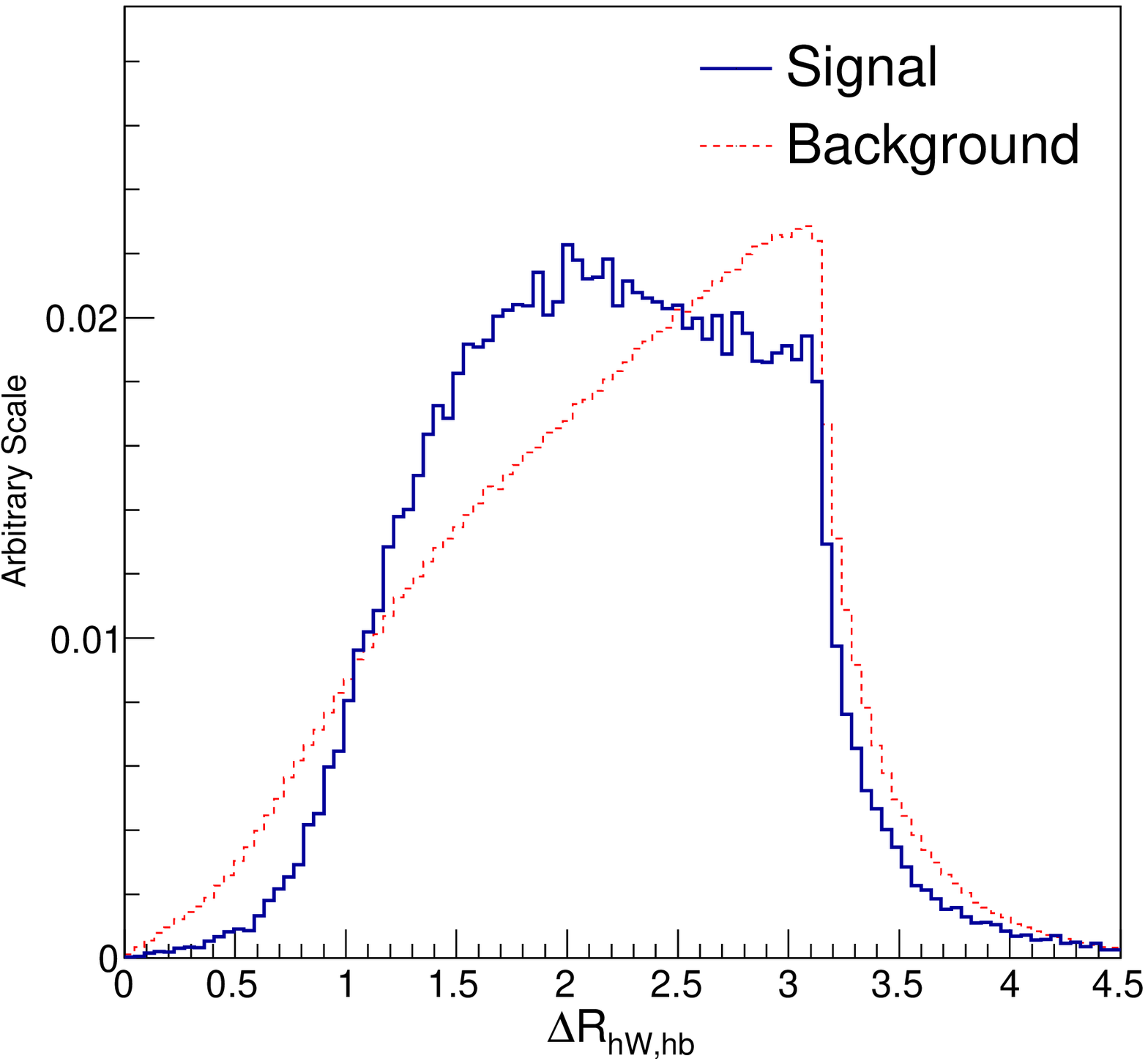}\\
\includegraphics[width=0.32\textwidth]{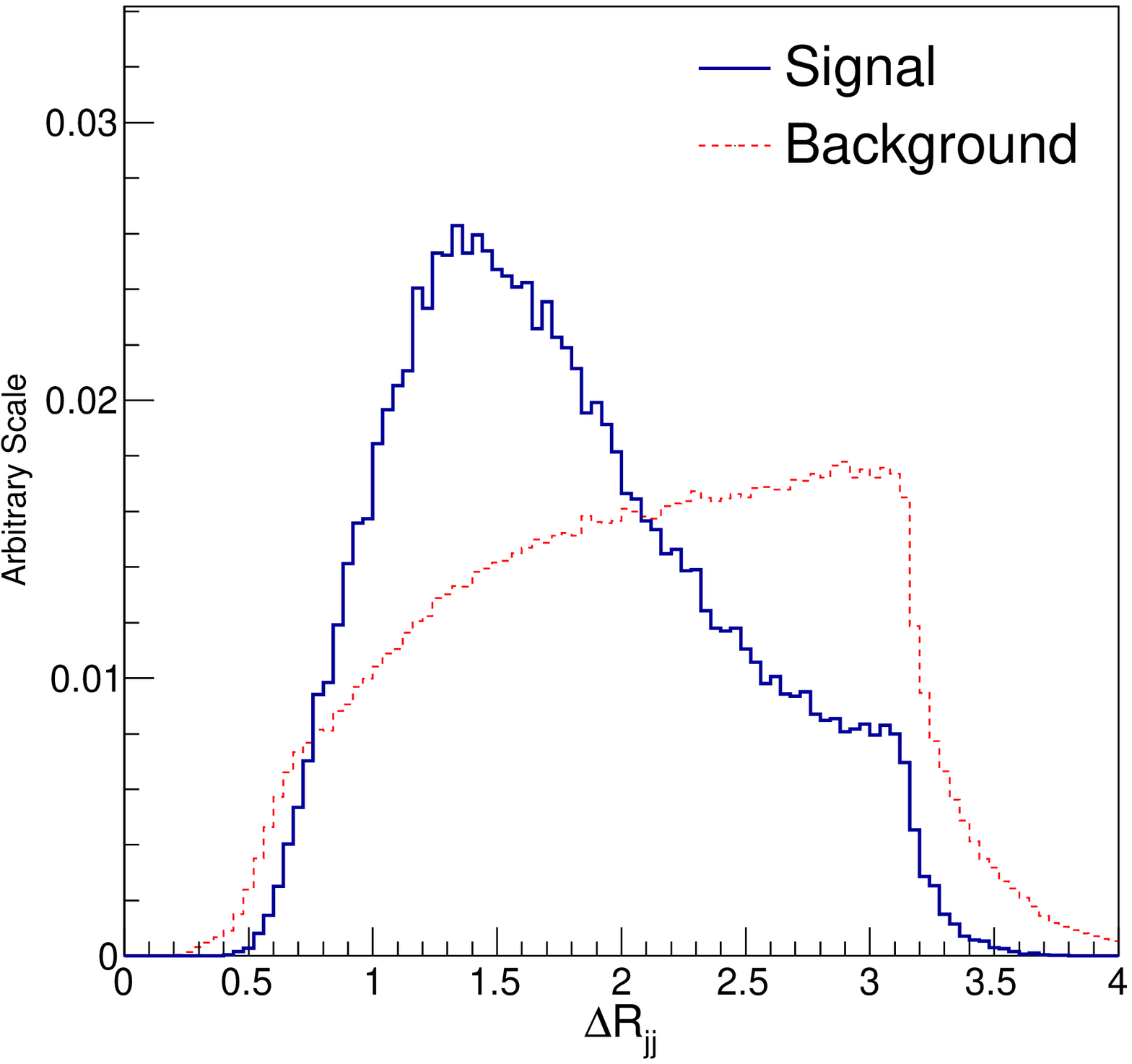}&
\includegraphics[width=0.32\textwidth]{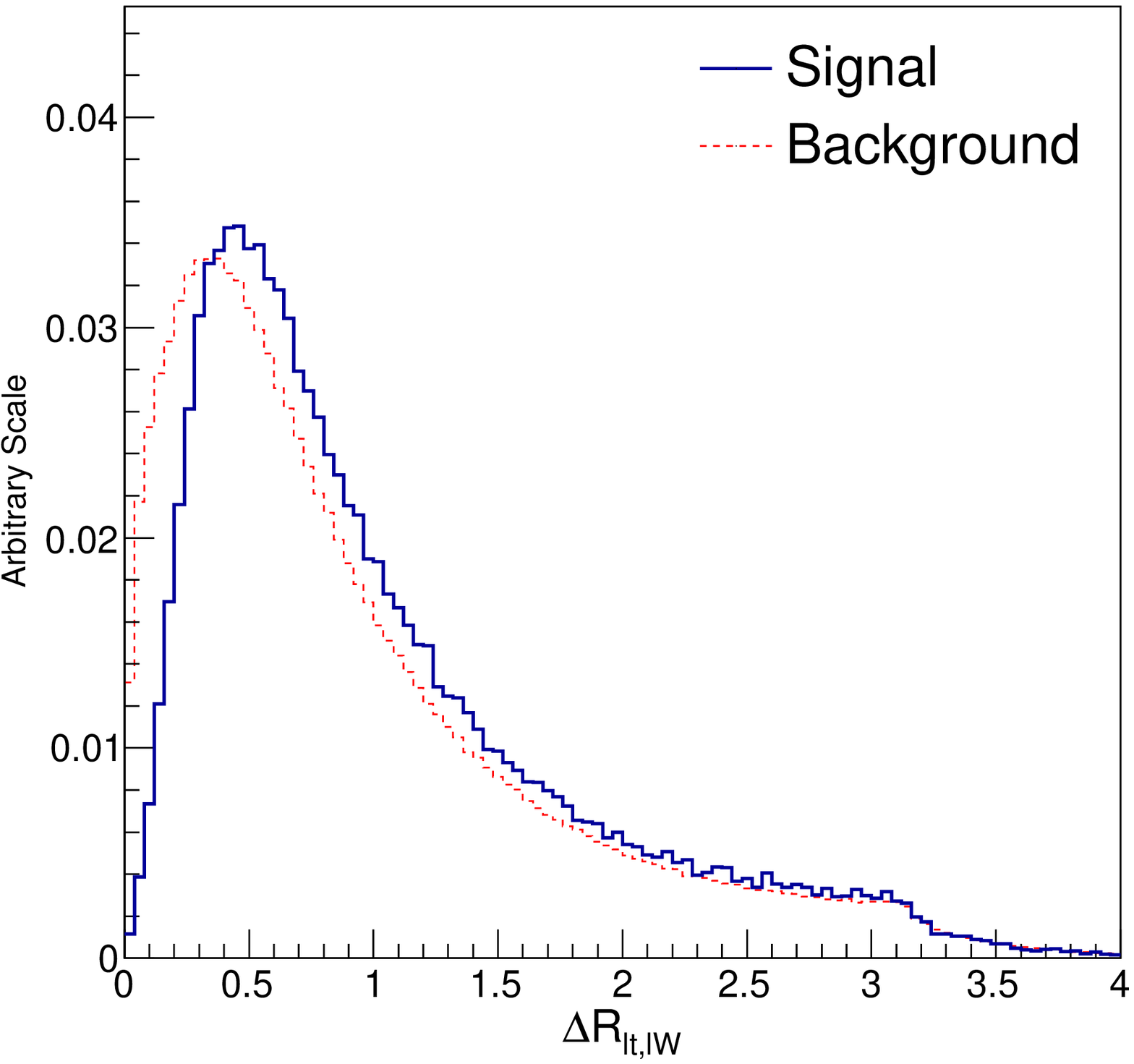}&
\includegraphics[width=0.32\textwidth]{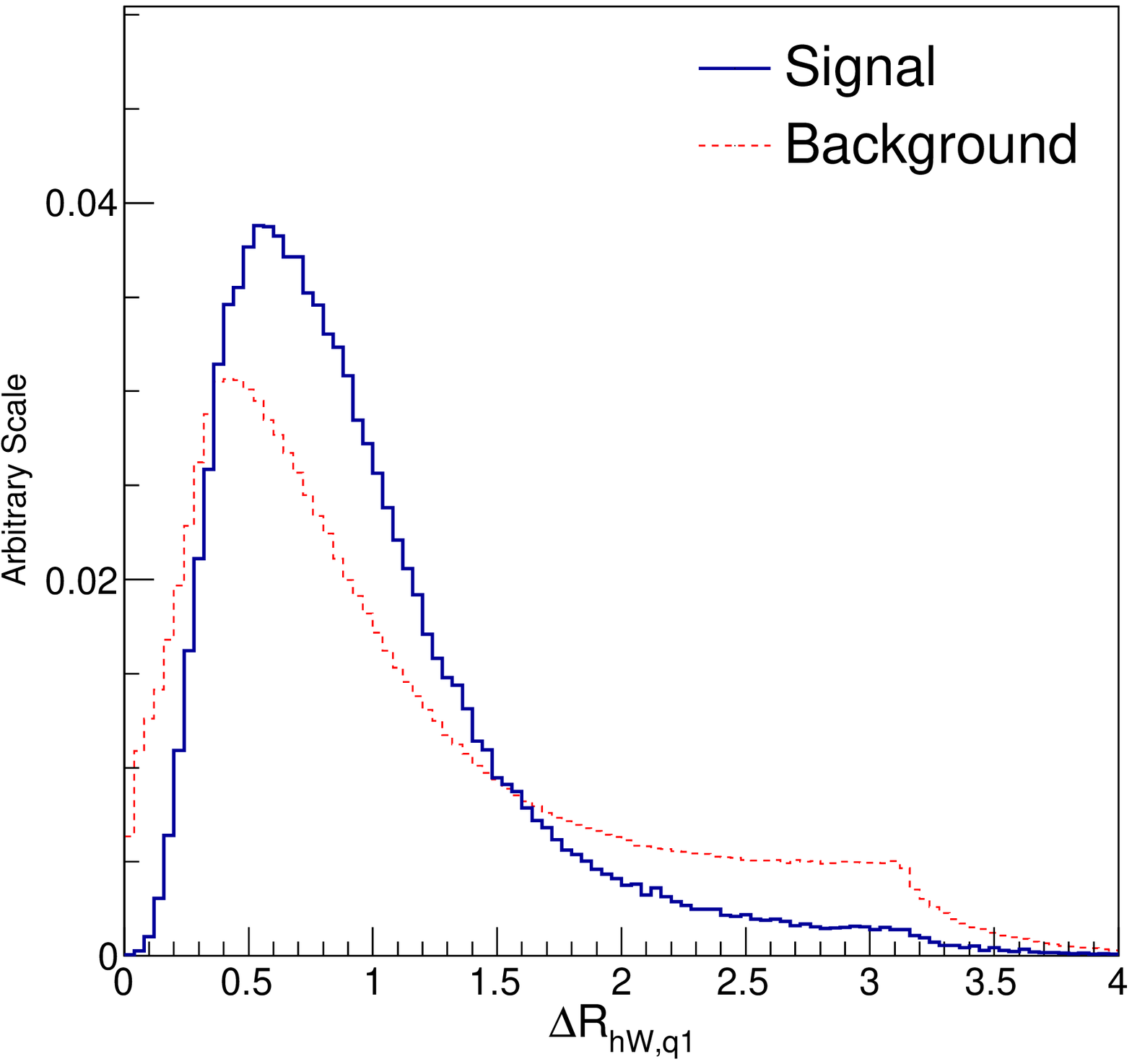}\\
\includegraphics[width=0.32\textwidth]{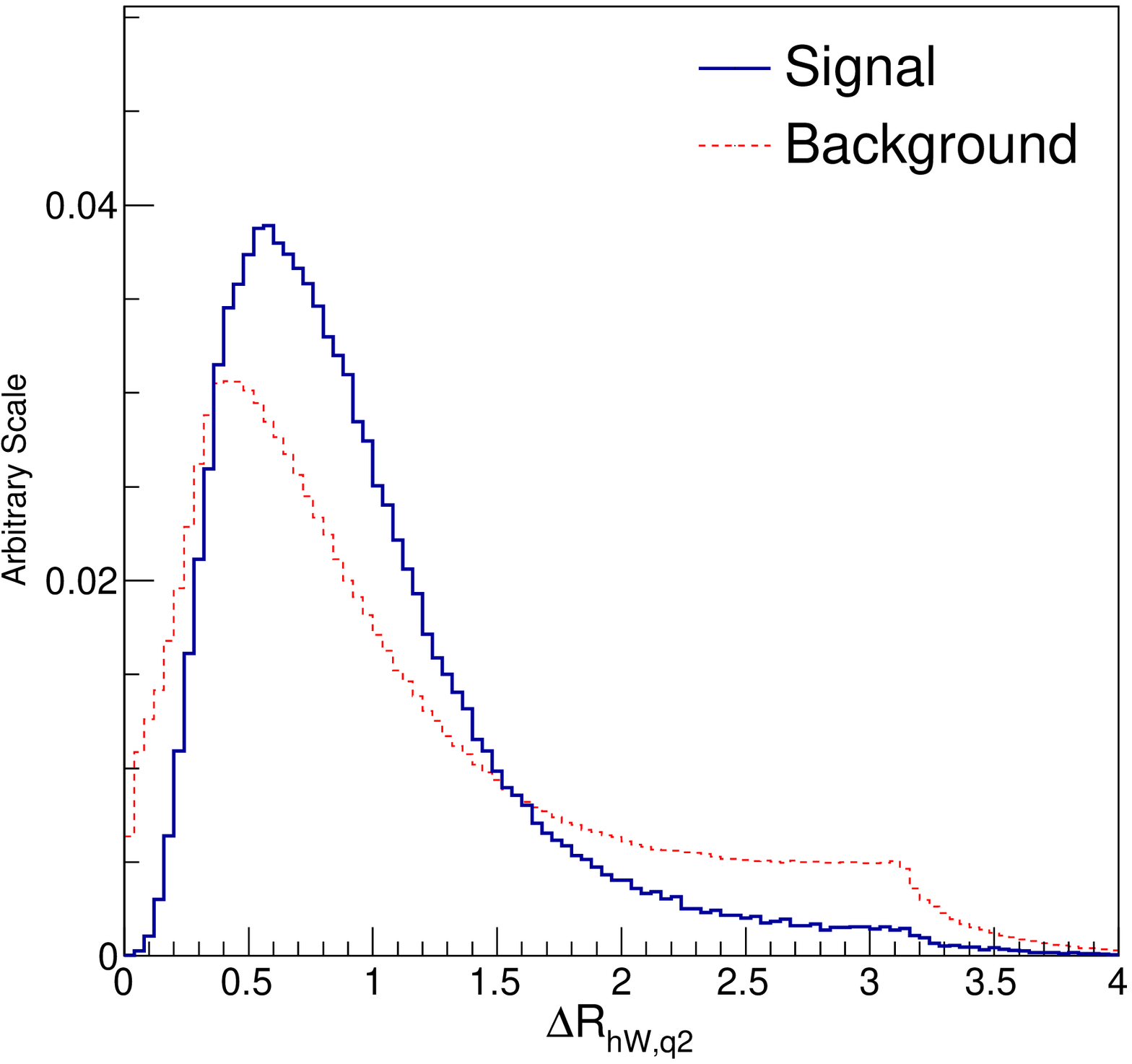}&
\includegraphics[width=0.32\textwidth]{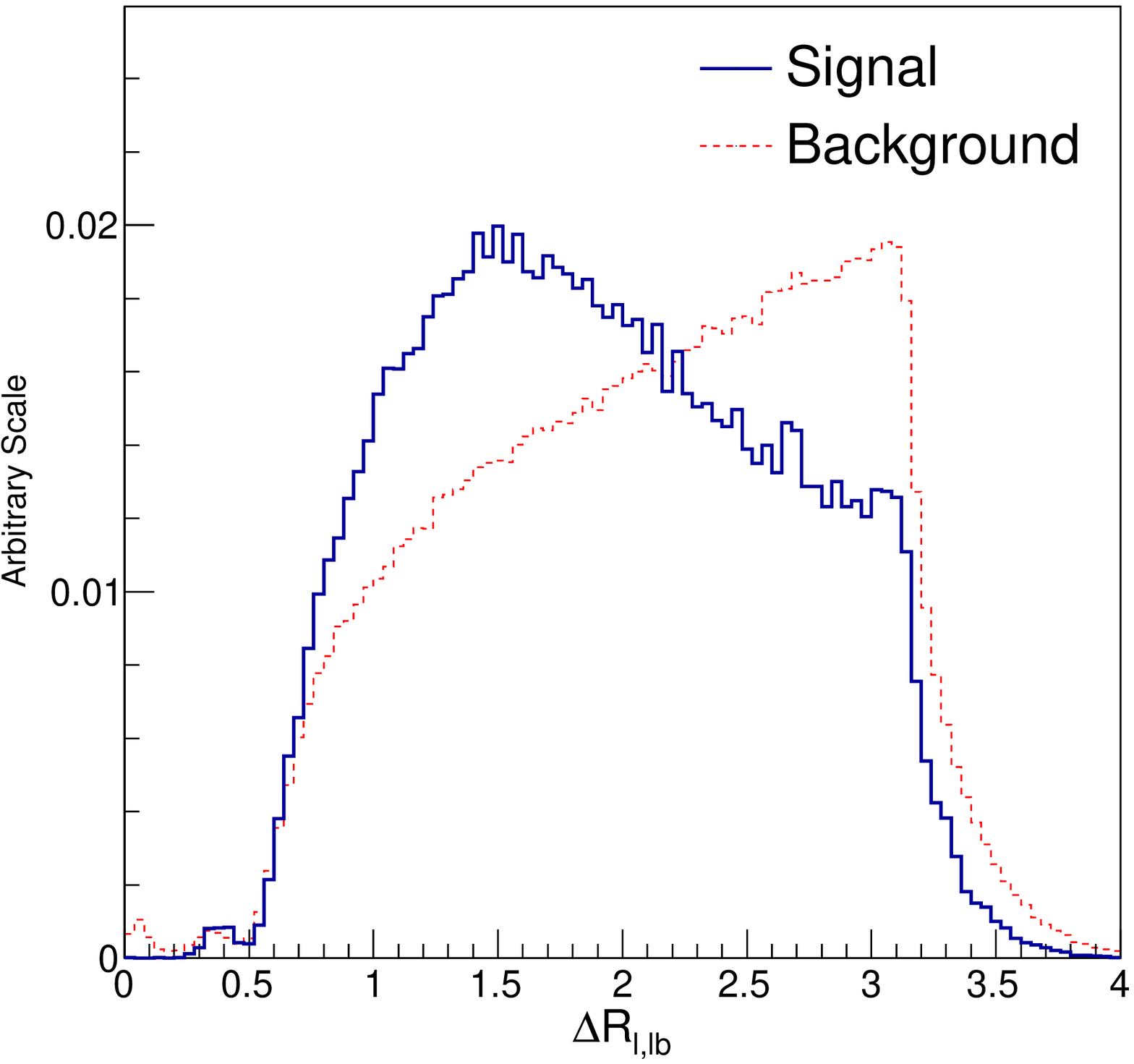}&
\includegraphics[width=0.32\textwidth]{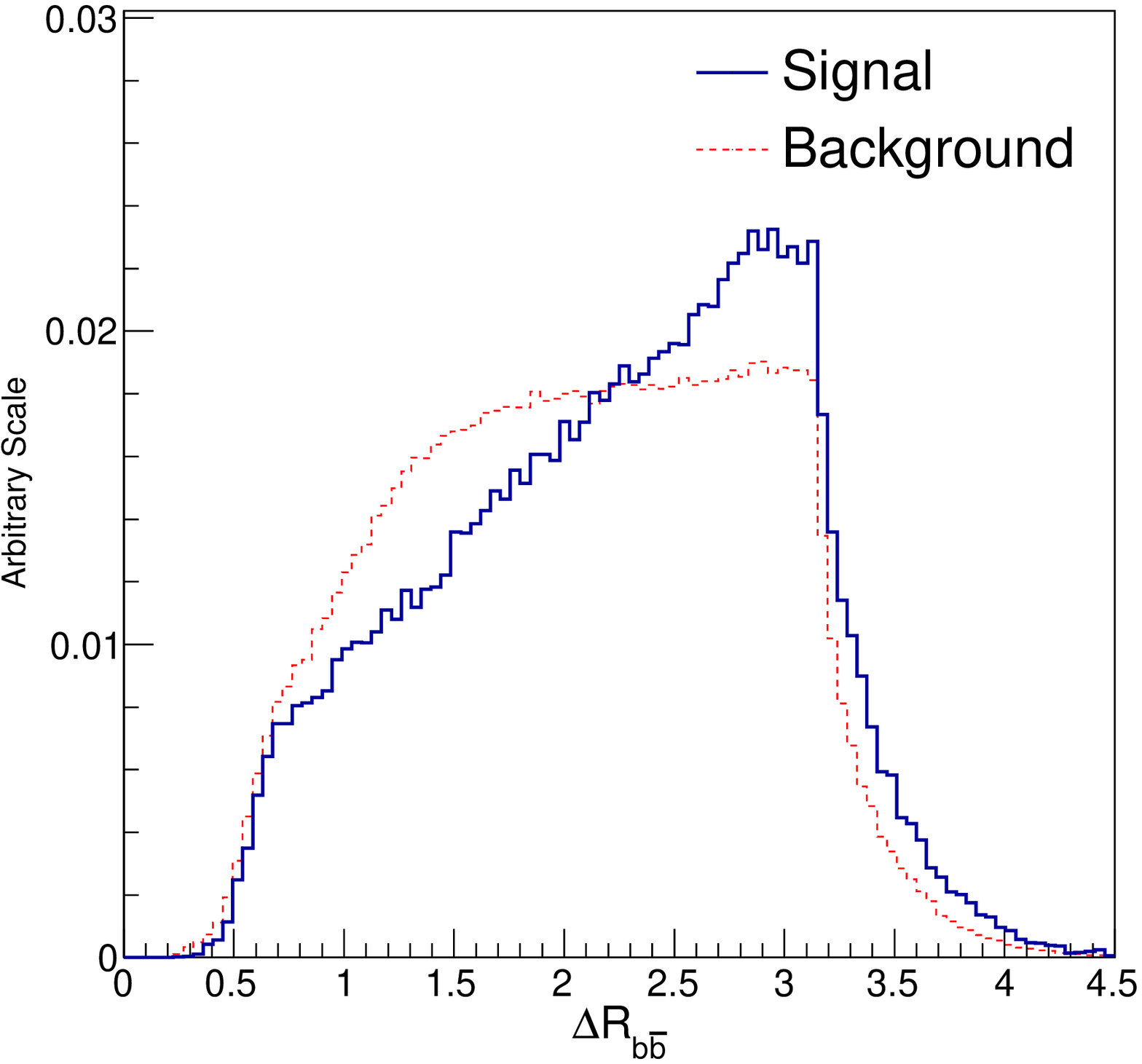}
\end{tabular}
\caption[Data fit]{Distributions of the neural network input variables for the \mtop measurement 
of the signal (all jets are correctly matched) and the background (at least one jet unmatched) 
using a SM \ttbar sample with $\mtop=\gevcc{173}$. 
}
\label{NN_mass}
\end{cfigure1c}

\begin{cfigure1c}
\begin{tabular}{cc}
\includegraphics[width=0.53\textwidth]{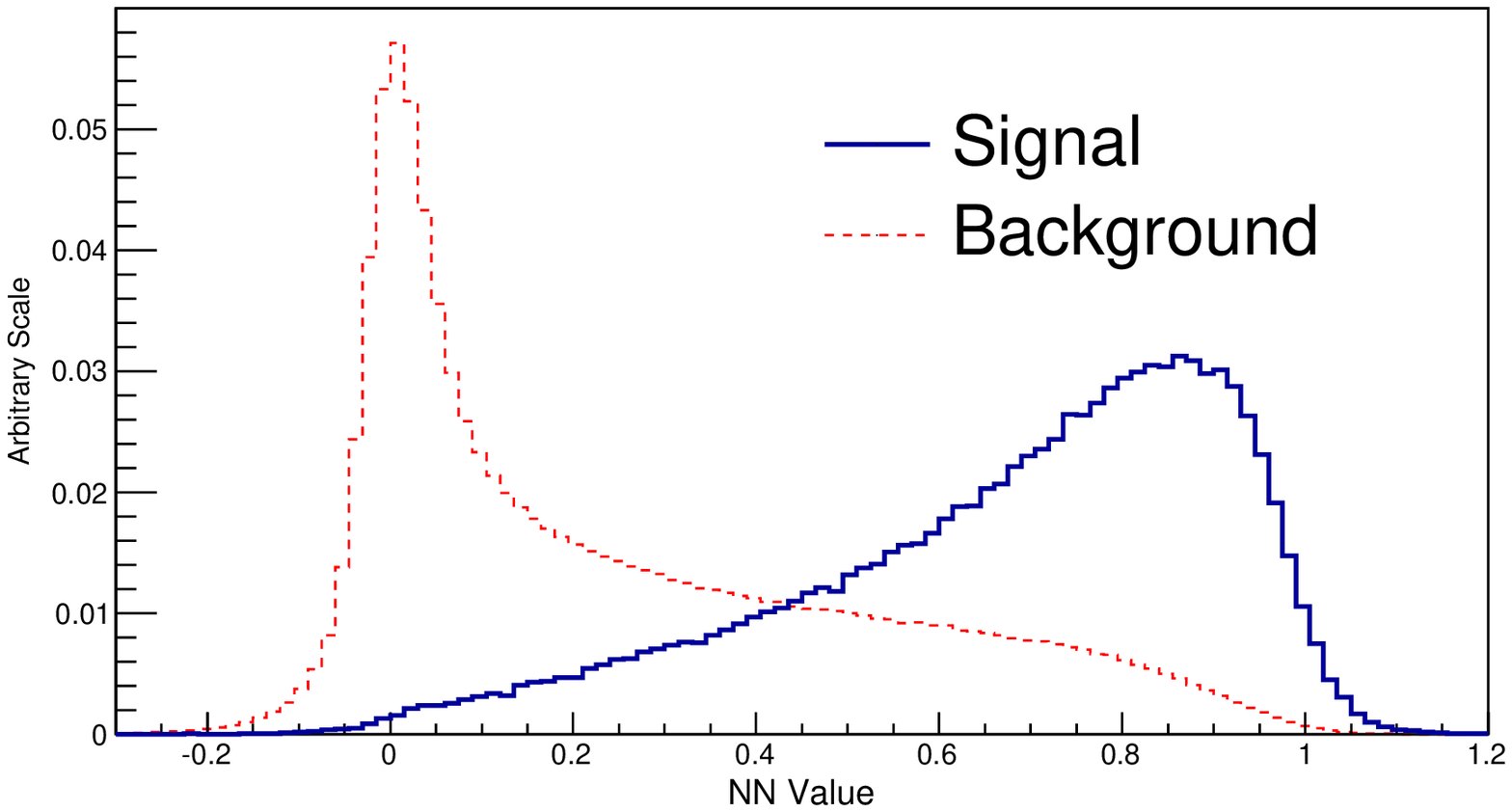}&
\includegraphics[width=0.40\textwidth]{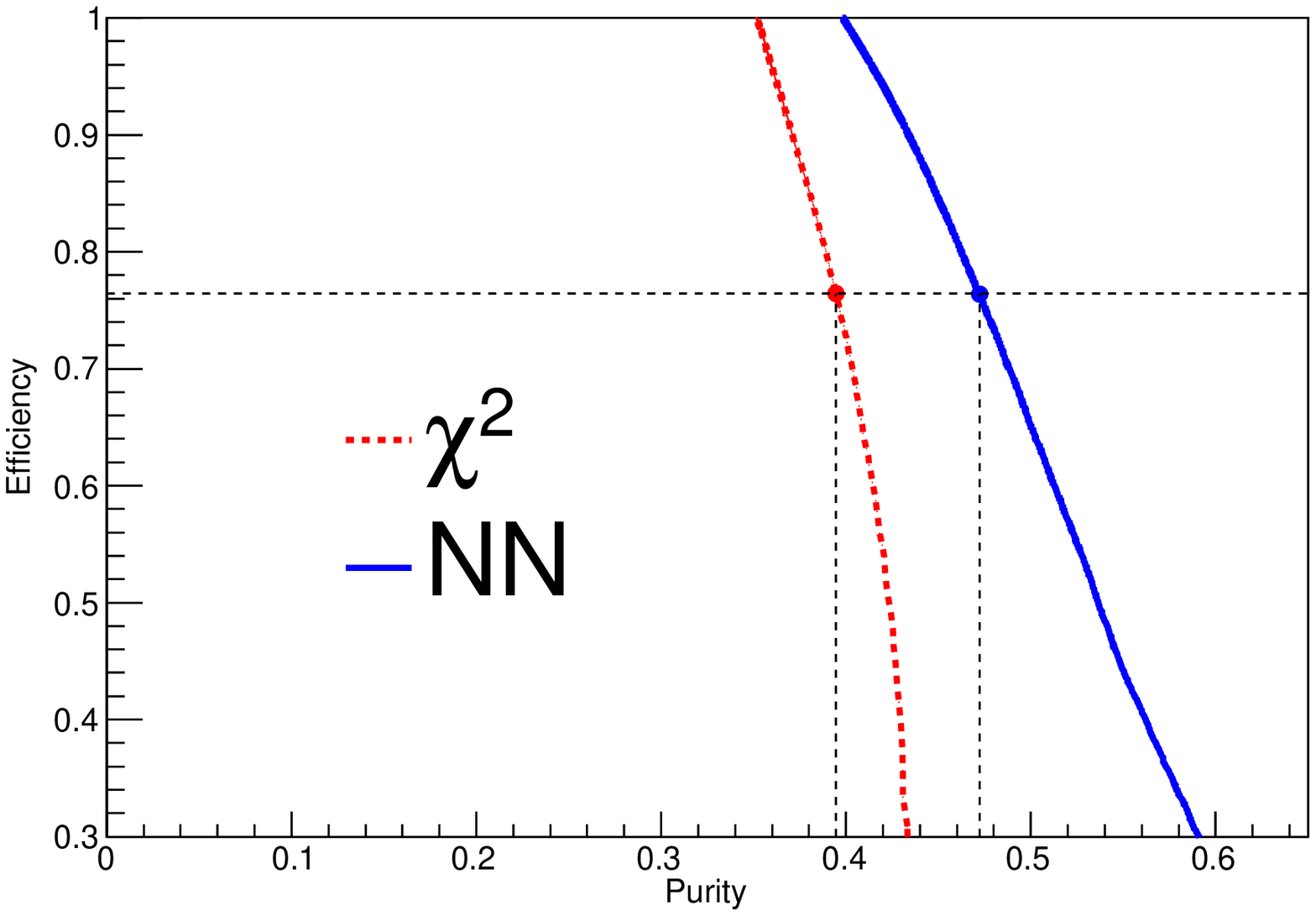}\\
(a) & (b) \\
\end{tabular}
\caption[Data fit]{(a) Distribution of the neural network output for the \mtop measurement of the 
signal and the background using a SM $\ttbar$ sample with $\mtop=\gevcc{173}$. (b) The relationship 
between the purity and efficiency for the NN method and 
the $\chi^2$ method is shown. At the same efficiency (76\%), which corresponds to the $\chi^2 <9$ 
cut, the NN method has 49\% purity, while the $\chi^2$ method has 39\% purity.
}
\label{NN_mass_out}
\end{cfigure1c}

For the \mtop measurement, we consider kinematic variables that do not depend on \mtop but
have a discriminant power between the signal and the background. In addition to the $\chi^2$ 
information, we consider the transverse momentum of the \ttbar system~$p_{T}^{t\bar{t}}$ as well 
as the azimuthal angle between the $t$ and $\bar{t}$ quarks~$\Delta \phi$~\cite{azidecorr}. 
Without additional radiation, $p_{T}^{t\bar{t}}$ and $\Delta \phi$ are predicted to be zero 
and $\pi$, respectively, because no transverse momentum exists at the initial collision. 
If these values are far from the predicted values, this may indicate an incorrect matching 
choice that is caused by combinatoric ambiguities.  
We also consider the distances~($\Delta R$) between the reconstructed particles, which are
useful for determining the correct matching choice. The NN training input variables are 
listed below:
\begin{enumerate}
\item  $\chi^2$: $\chi^2$ value of the kinematic reconstruction.
\item $p_{T}^{t\bar{t}}$: Transverse momentum of the \ttbar system.
\item $\Delta \phi$: Azimuthal angle between the $t$ and $\bar{t}$ quarks.
\item $\Delta R$: Distance between the $t$ and $\bar{t}$ quarks.
\item $\Delta R_{lW,lb}$: Distance between the $b$ quark and $W$ boson of the 
leptonically decaying $t$ quark.
\item $\Delta R_{hW,hb}$: Distance between the $b$ quark and $W$ boson of the 
hadronically decaying $t$ quark.
\item $\Delta R_{jj}$: Distance between the two jets of the hadronically decaying $W$ boson.
\item $\Delta R_{lt,lW}$: Distance between the leptonically decaying $t$ quark and its 
decaying daughter $W$ boson.
\item $\Delta R_{ht,hW}$: Distance between the hadronically decaying $t$ quark and its 
decaying daughter $W$ boson.
\item $\Delta R_{hW,q1}$: Distance between the hadronically decaying $W$ boson and its 
decaying daughter quark(1).
\item $\Delta R_{hW,q2}$: Distance between the hadronically decaying $W$ boson and its 
decaying daughter quark(2).
\item $\Delta R_{l,lb}$: Distance between the lepton and $b$ quark of the leptonically
decaying $t$ quark.
\item $\Delta R_{b\bar{b}}$: Distance between the $b$ and $\bar{b}$ quark.
\end{enumerate}

\begin{cfigure1c}
\begin{tabular}{cc}
\includegraphics[width=0.35\textwidth]{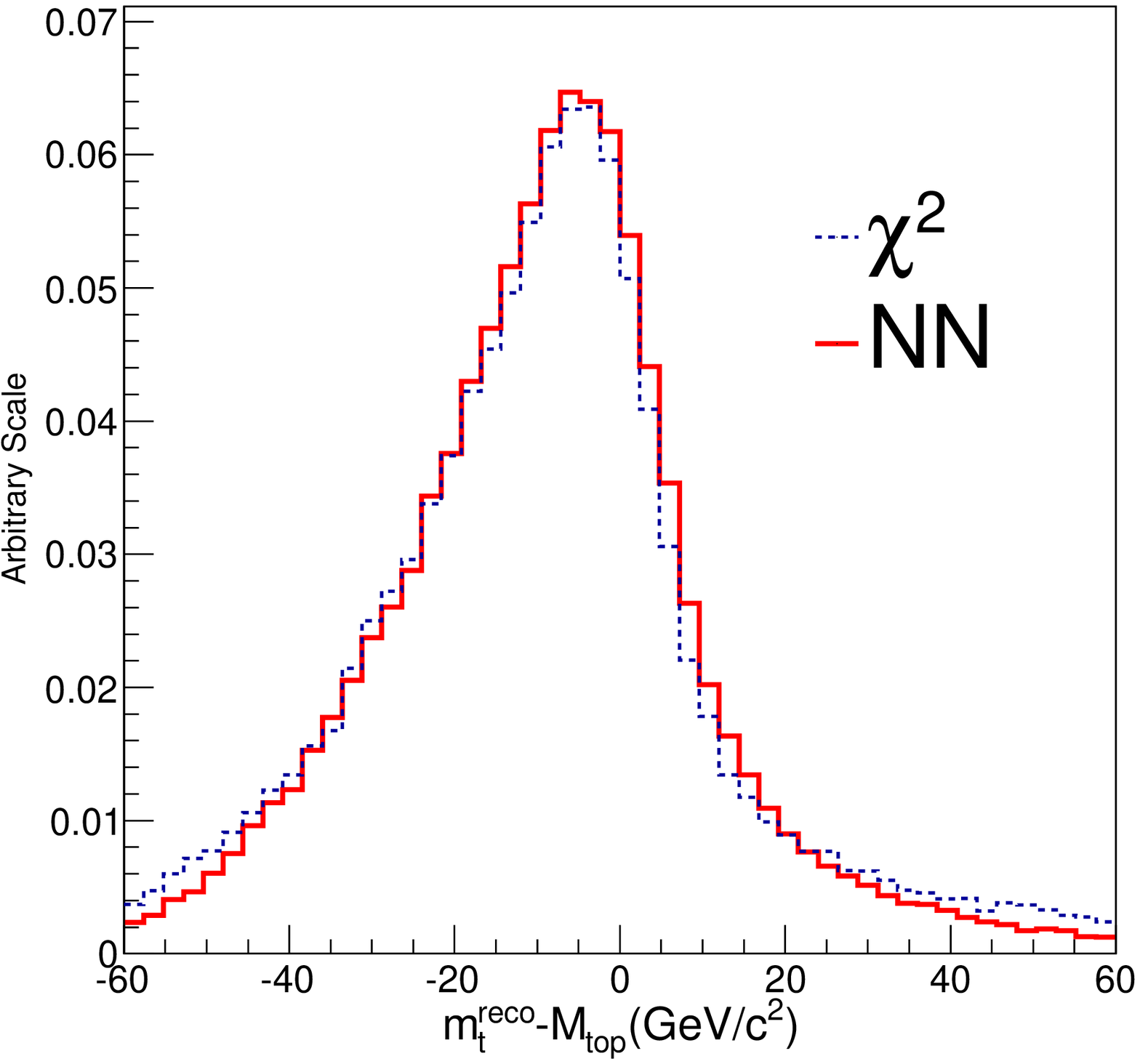}&
\includegraphics[width=0.55\textwidth]{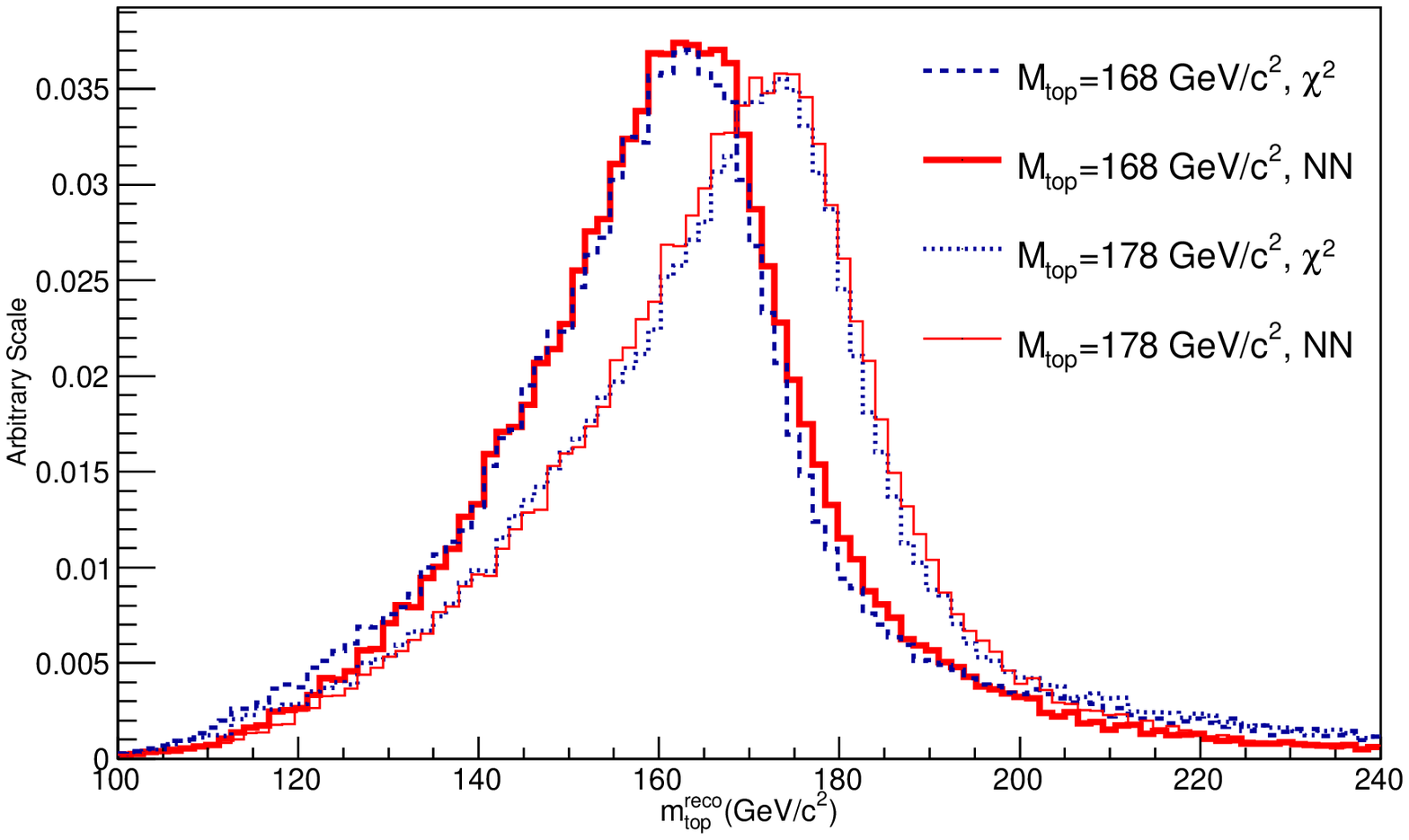}\\
(a) & (b) \\
\end{tabular}
\caption[Data fit]{ (a) The \mtreco minus \mtop distributions of the \ttbar sample using NN 
method~(solid line) and the $\chi^2$ method~(dashed line) with \mtop=\gevcc{173} are shown. 
The NN method has Better resolution (close to zero). (b)  The \mtreco distributions of the 
\gevcc{168} \ttbar sample using the NN method~(thick solid line) and the $\chi^2$ 
method~(thick dotted line), and the \gevcc{178} \ttbar sample using the NN method~(thin solid 
line) and the $\chi^2$ method~(thin dotted line) are presented. The NN method has better 
separation between the \gevcc{168} and \gevcc{178} samples. 
}
\label{NN_mass_comp}
\end{cfigure1c}

The distributions and the separation power of the input variables used in the NN for both the 
signal and the background are shown in Fig.~\ref{NN_mass}. Even though $\chi^2$ is useful 
for determining the correct matching, all of the other variables also provide meaningful 
discrimination between the signal and the background. Our NN configuration has twelve input 
variables, two hidden layers, and one output node. After training, we process the SM \ttbar 
sample using the trained NN. Figure~\ref{NN_mass_out} (a) shows the NN output value~(NN$_{\text{out}}$) 
of the signal and the background. We find that NN$_{\text{out}}$ produces a good separation 
power between the signal and the background. 

In the $\chi^2$ method, a candidate for the correct matching combination is chosen by the 
case that has the minimum $\chi^2$ value. However, in the NN method, we choose the combination 
with the maximum NN$_{\text{out}}$. Because CDF analysis usually rejects poorly reconstructed 
events by requiring $\chi^2 <9$, we also try to remove poorly reconstruction events in the 
NN method using maximum NN$_{\text{out}}$ requirements. The purity of an event reconstruction 
is highly dependent on a fraction of the event passing the criteria (efficiency). We therefore 
study the relationship  between the purity and the efficiency using each reconstruction method. 
Figure~\ref{NN_mass_out} (b) shows the efficiency as a function of the purity for both the $\chi^2$ 
method and the NN method. As we can see, the NN method has much higher purity for the same 
efficiency.  If we select 76\% efficiency in the NN method, which corresponds to an efficiency 
of $\chi^2<9$ in the $\chi^2$ method, the NN$_{\text{out}}$ criteria should be 
NN$_{\text{out}}>0.60$. With this condition, the NN method has 47\% purity, which is 
approximately 21\% better than the $\chi^2$ method.  

For realistic \mtop measurements, we study the reconstructed top-quark mass~(\mtreco) distribution, 
which is an observable of the \mtop measurement~\cite{cdf_fitter}.  We first examine the 
difference between the \mtreco and the true \mtop value of the SM \ttbar sample. 
Figure~\ref{NN_mass_comp}~(a) shows the distributions of \mtreco minus \mtop using two 
different reconstruction methods. As we can see, the NN method has better resolution (~10\%) 
than the $\chi^2$ method. To study any bias on \mtop in the NN reconstruction, we generate 
two additional SM \ttbar samples that have different \mtop values (\gevcc{168} and \gevcc{178}).  
We apply the NN trained by the \mtop=\gevcc{173} sample to both samples and compare the 
\mtreco distributions. As shown in Fig.~\ref{NN_mass_comp}~(b), the \mtreco distribution is 
slightly smaller using the NN method. However, the mean value changes from the different mass 
samples are quite similar between the $\chi^2$ method and the NN method. To quantify the 
performance, we calculate $\Delta \mtreco (\gevcc{178}-\gevcc{168})/\text{RMS}$, where 
\text{RMS} is average root-mean-square of the \mtreco distributions. We obtain an approximately 
11\% higher value using the NN method than that of $\chi^2$ method. Therefore, we can achieve 
better precision of the \mtop measurement using the NN method with a quantitatively similar 
improvement in the statistical uncertainty.  

\section{\afb measurement} 
For the $\afb$ measurement, we consider kinematic variables that do not depend on the $\theta$ 
angle. We therefore do not consider the angles between particles. Because \mtop is very 
precisely measured~\cite{tevave,lhcave}, we assume \mtop=\gevcc{173} is appropriate for 
\afb measurements. With this assumption, the reconstructed masses of the particles are useful 
for denoting the correct combination. In general, incorrect combinations will yield smaller 
reconstructed masses as well as lower resolutions. The full list of input variables for 
the \afb measurement are shown below:
\begin{enumerate}
\item  $\chi^2$: $\chi^2$ value of the kinematic reconstruction.
\item $p_{T}^{t\bar{t}}$: Transverse momentum of the \ttbar system.
\item $m^{t\bar{t}}$: Reconstructed invariant mass of the \ttbar system.
\item  $\mtreco$: Reconstructed top-quark mass.
\item  $m_{lt}$: Reconstructed mass of the leptonically decaying $t$ quark.
\item  $m_{ht}$: Reconstructed mass of the hadronically decaying $t$ quark.
\item  $m_{bl}$: Reconstructed invariant mass of the $b$ quark and lepton in the 
leptonically decaying $t$ quark.
\item  $m_{W}$: Reconstructed mass of the hadronically decaying $W$ boson.
\end{enumerate}

\begin{cfigure1c}
\begin{tabular}{ccc}
\includegraphics[width=0.32\textwidth]{V_chi2.eps}&
\includegraphics[width=0.32\textwidth]{V_pttbar.eps}&
\includegraphics[width=0.32\textwidth]{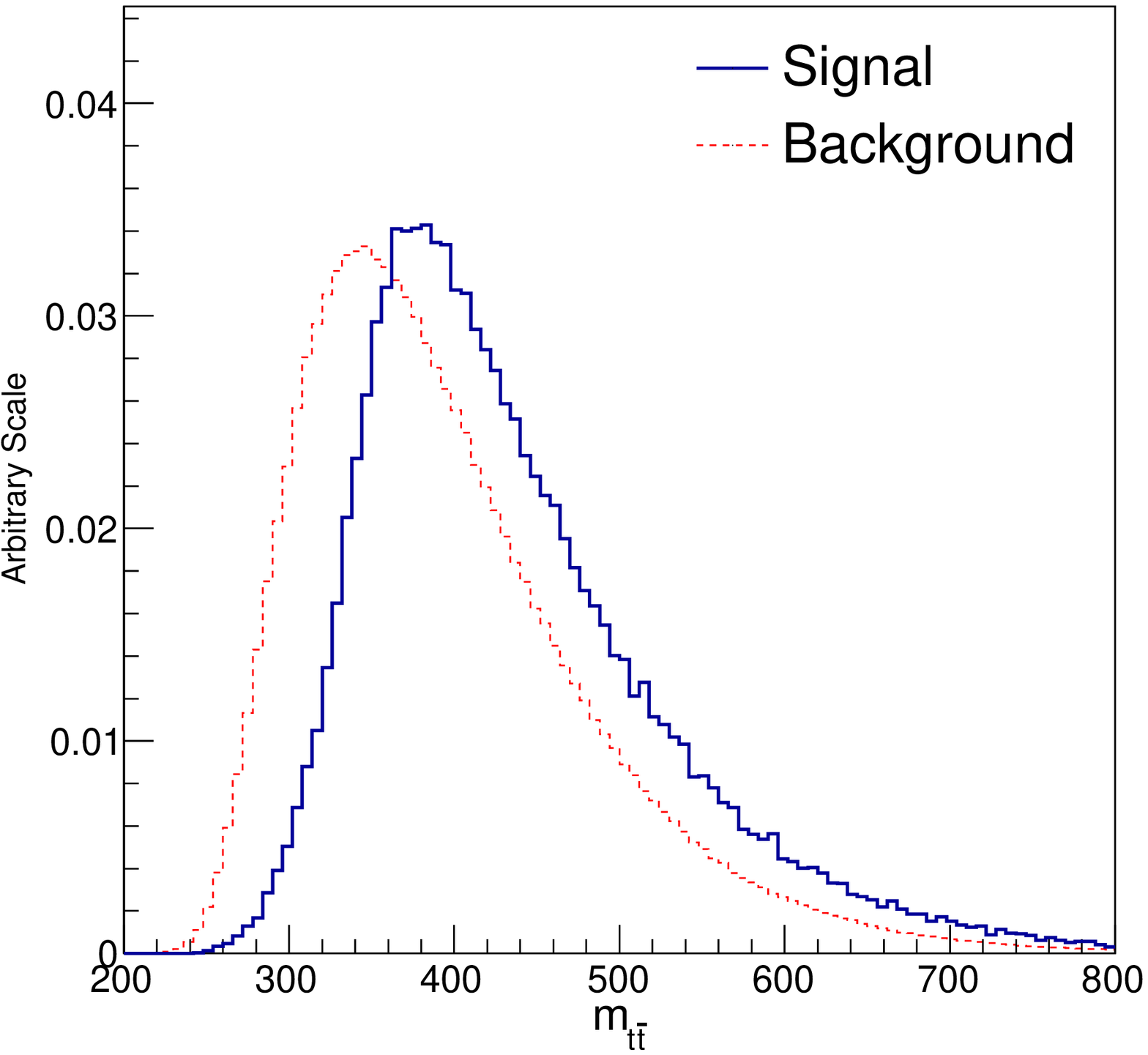}\\
\includegraphics[width=0.32\textwidth]{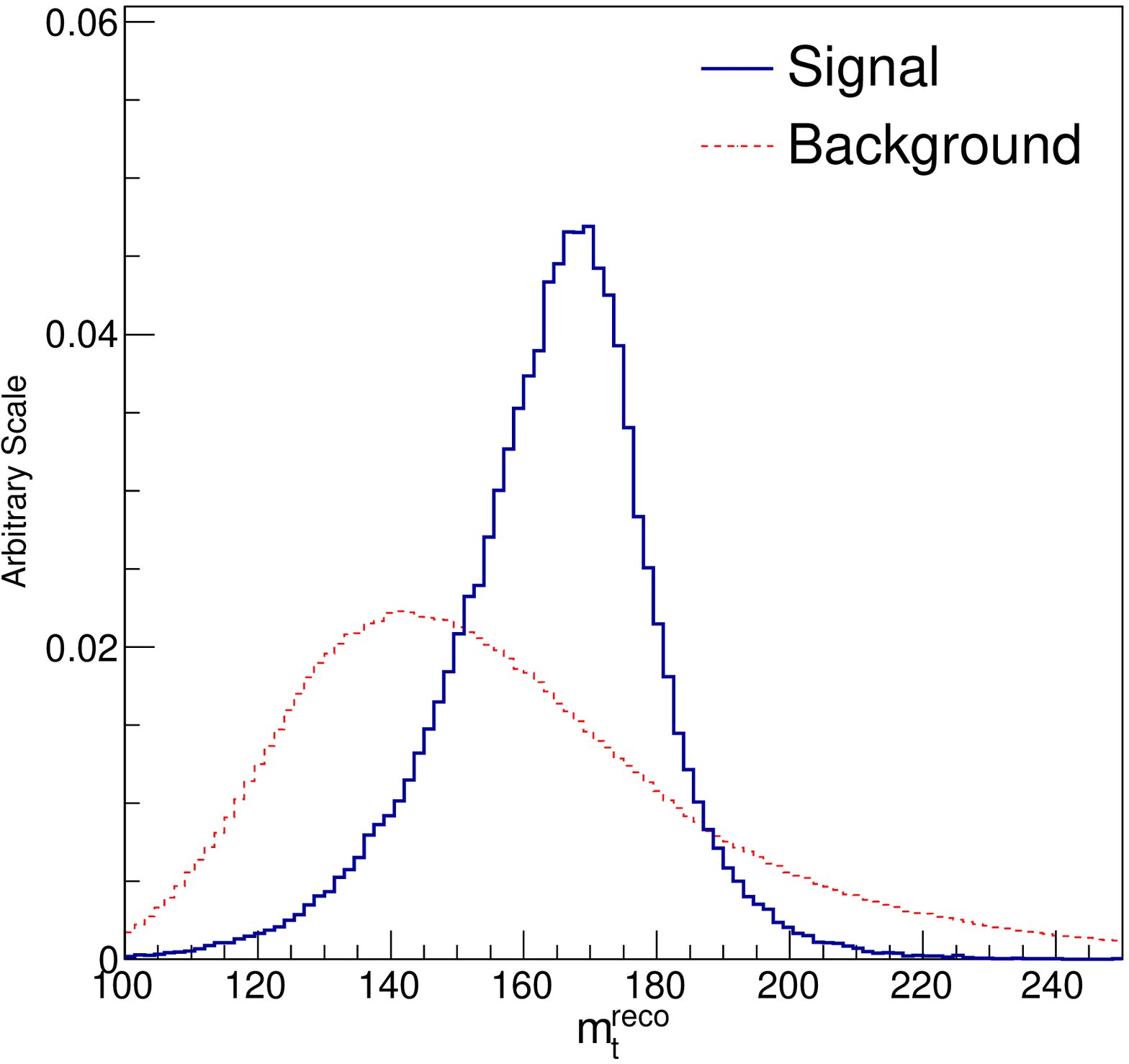}&
\includegraphics[width=0.32\textwidth]{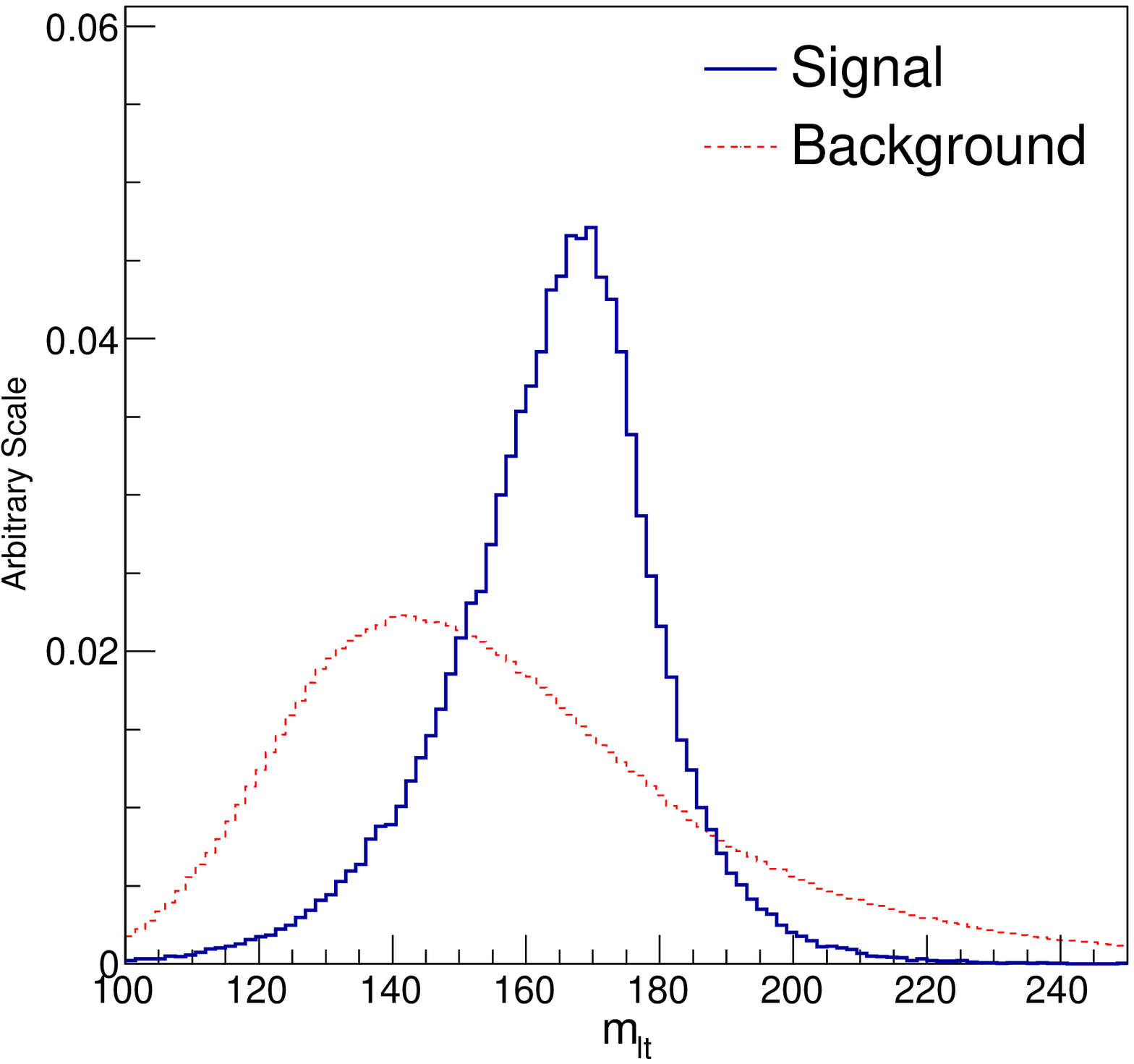}&
\includegraphics[width=0.32\textwidth]{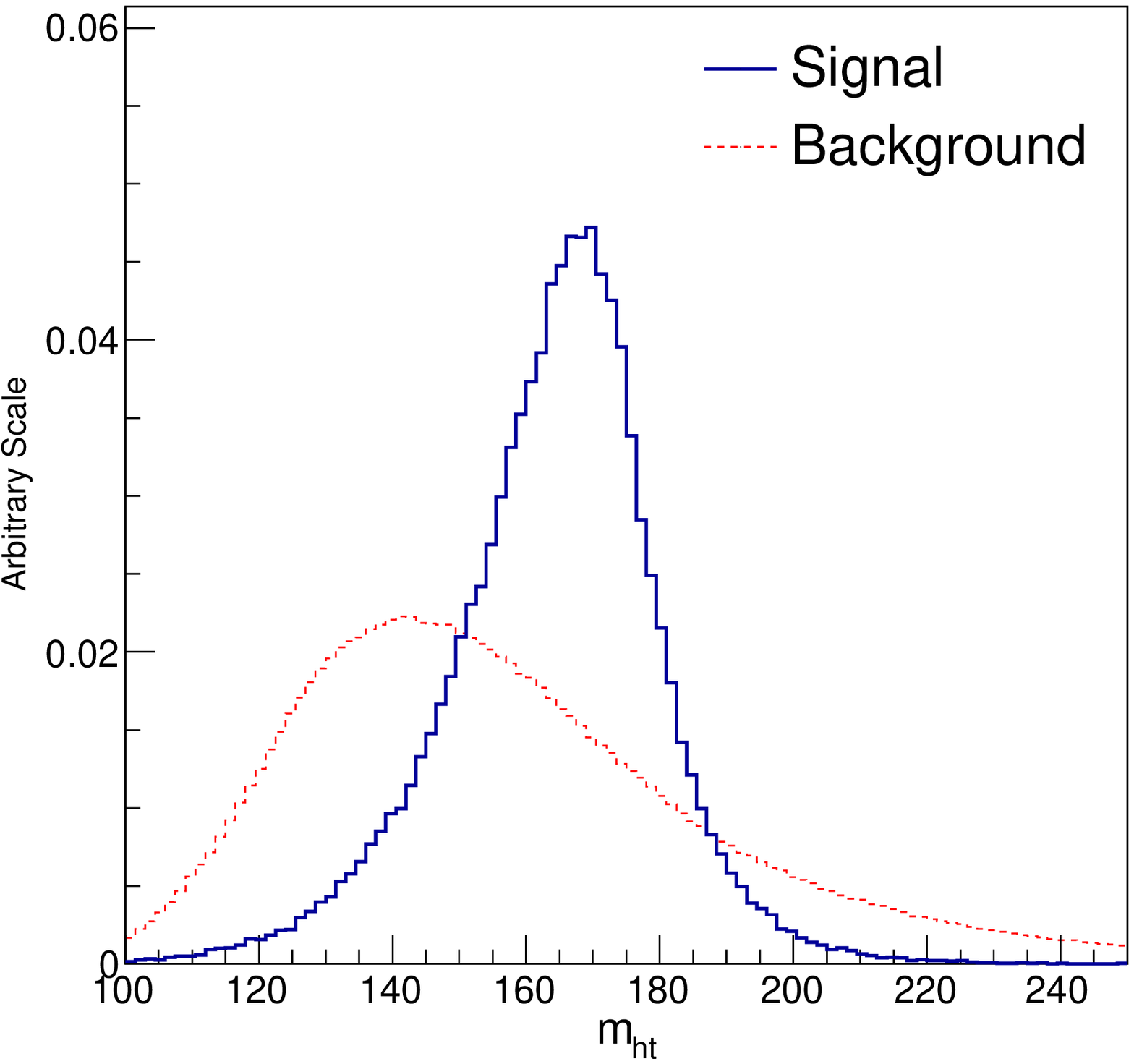}\\
\includegraphics[width=0.32\textwidth]{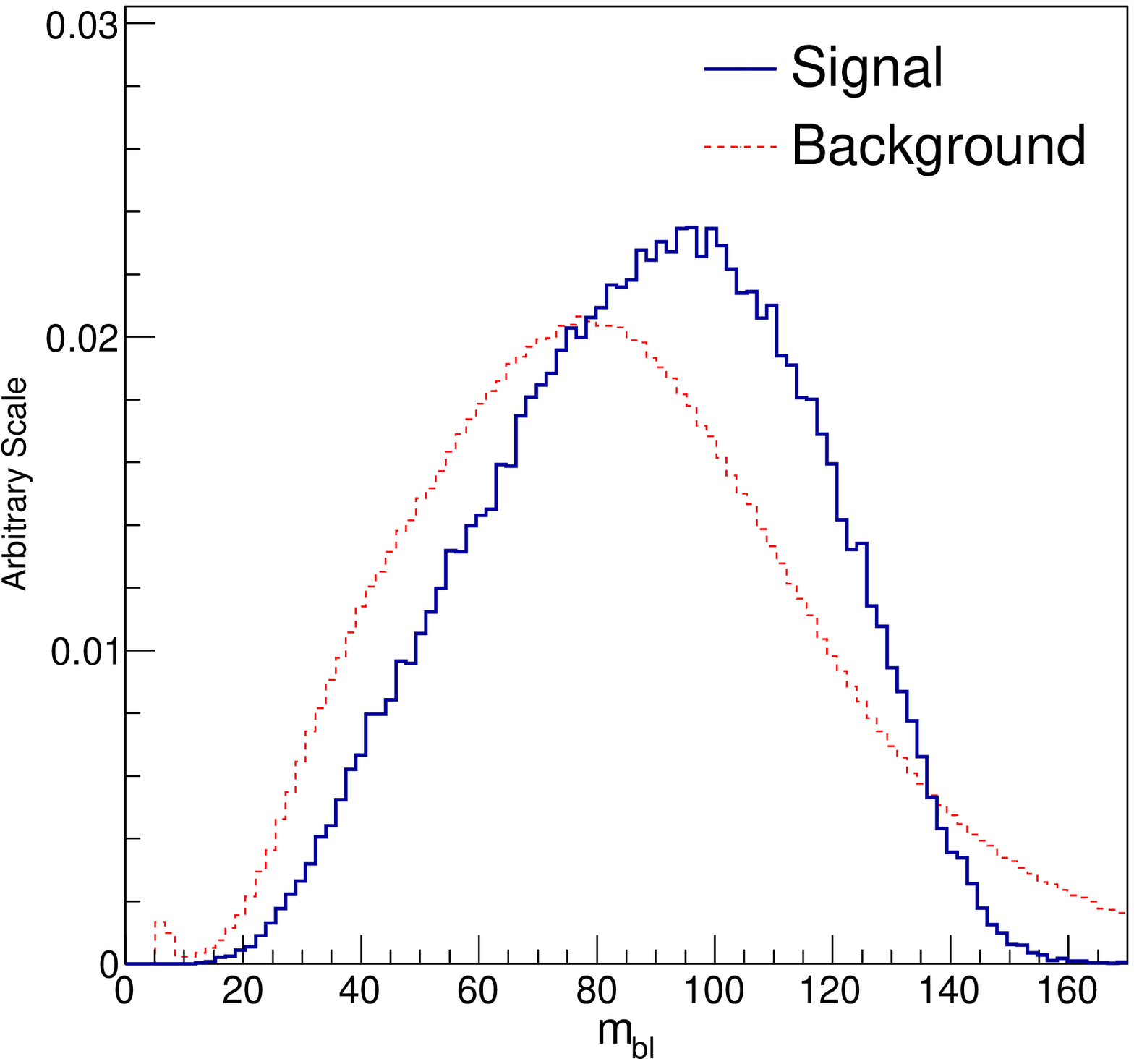}&
\includegraphics[width=0.32\textwidth]{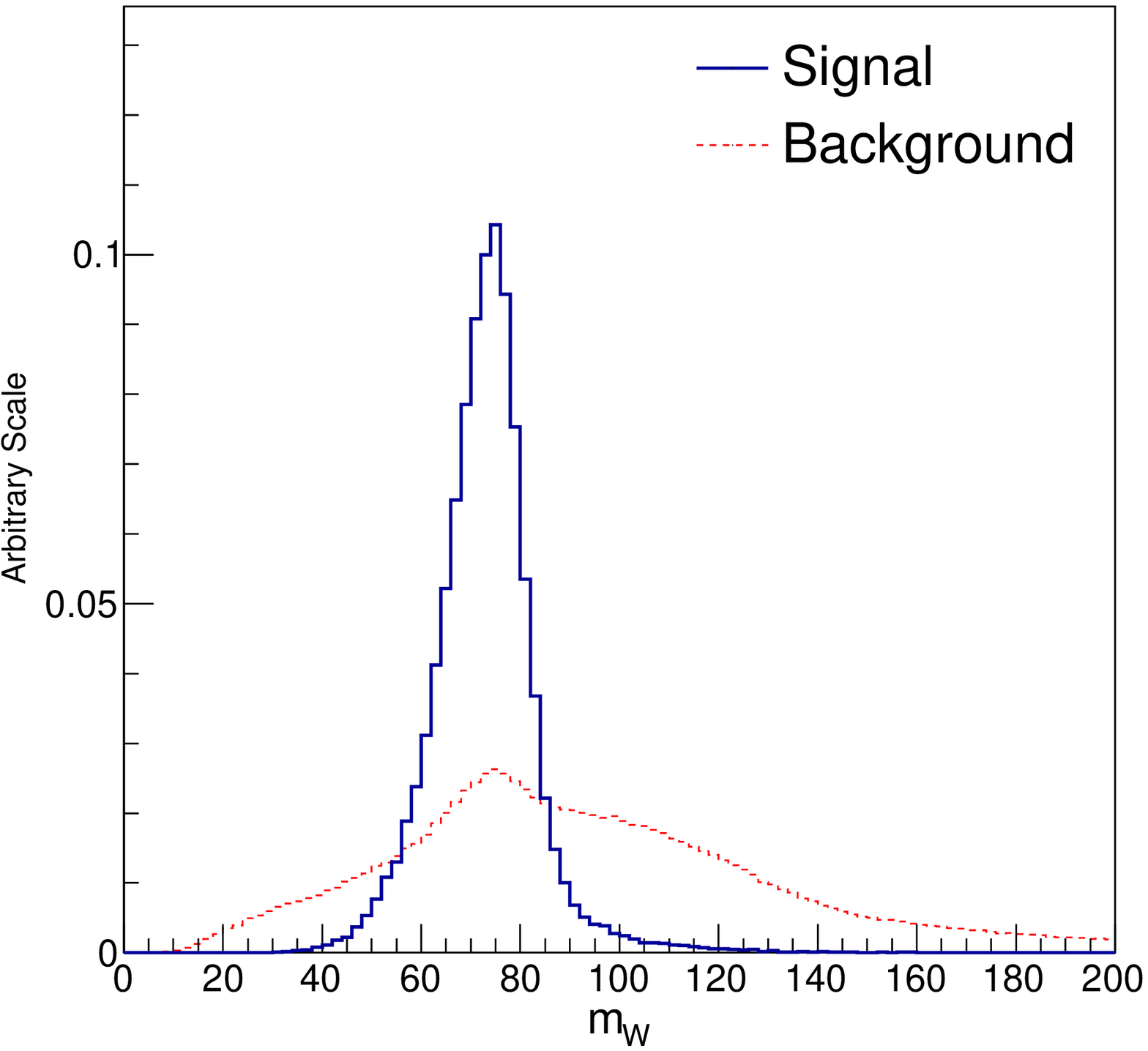}&
\end{tabular}
\caption[Data fit]{Distributions of the neural network input variables for the \afb measurement 
of the signal (all jets are correctly matched) and the background (at least one jet unmatched) 
using a SM \ttbar sample with $\mtop=\gevcc{173}$. 
}
\label{NN_afb}
\end{cfigure1c}

The distributions and separation power of the input variables used in the NN for both the signal 
and the background are shown in Fig.~\ref{NN_afb}. As we can see, the invariant masses of 
the reconstructed particles are very good discriminants. Our NN configuration for the \afb 
measurement has eight input variables, two hidden layers, and one output node. After training, 
we process the SM \ttbar sample with the trained NN. Figure~\ref{NN_afb_out} (a) shows 
NN$_{\text{out}}$ for the signal and the background using the SM sample. We achieve a very 
good separation between the signal and the background. 
We also apply the NN method to select the maximum NN$_{\text{out}}$ combination. 

\begin{cfigure1c}
\begin{tabular}{cc}
\includegraphics[width=0.53\textwidth]{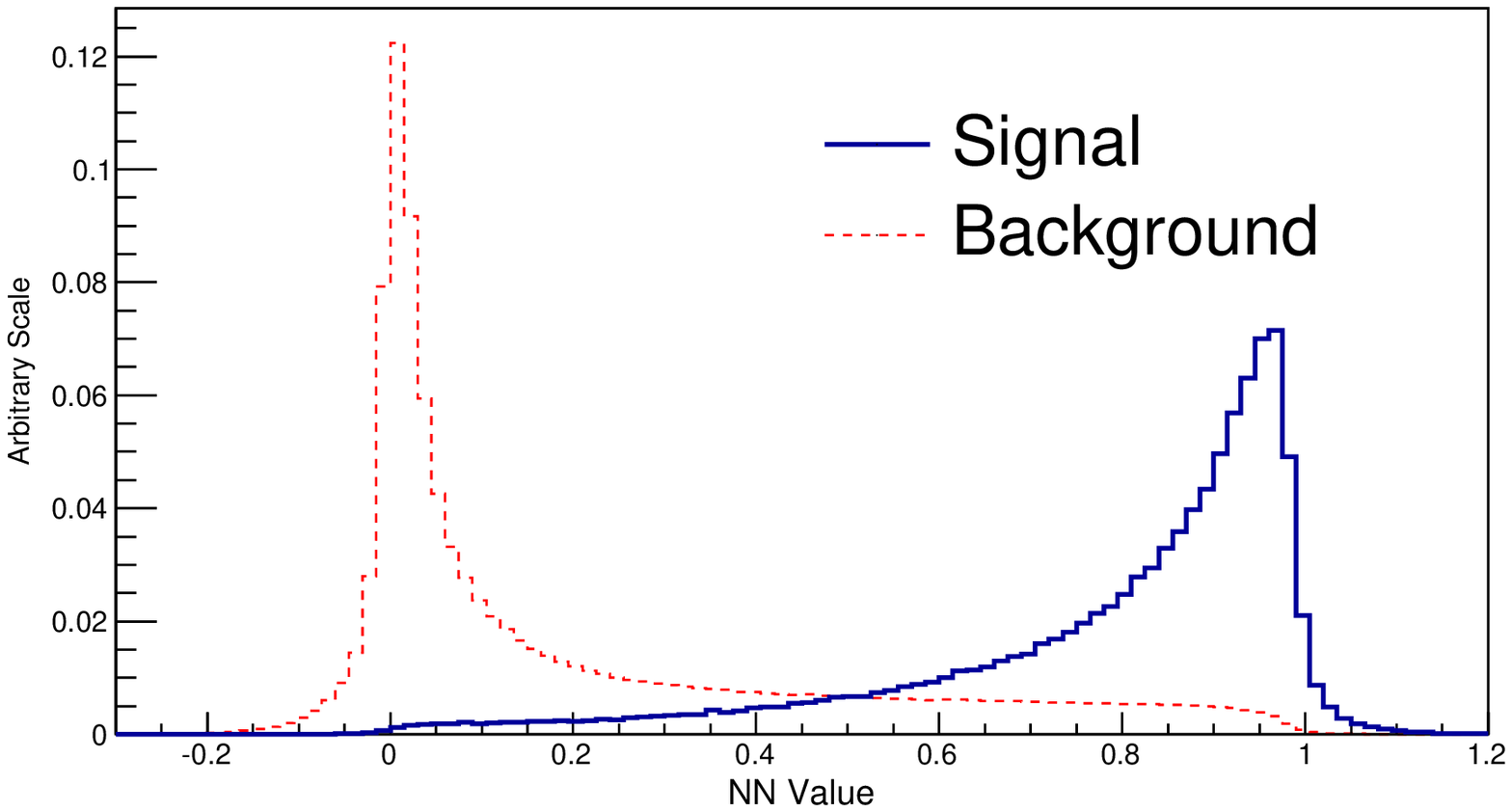}&
\includegraphics[width=0.40\textwidth]{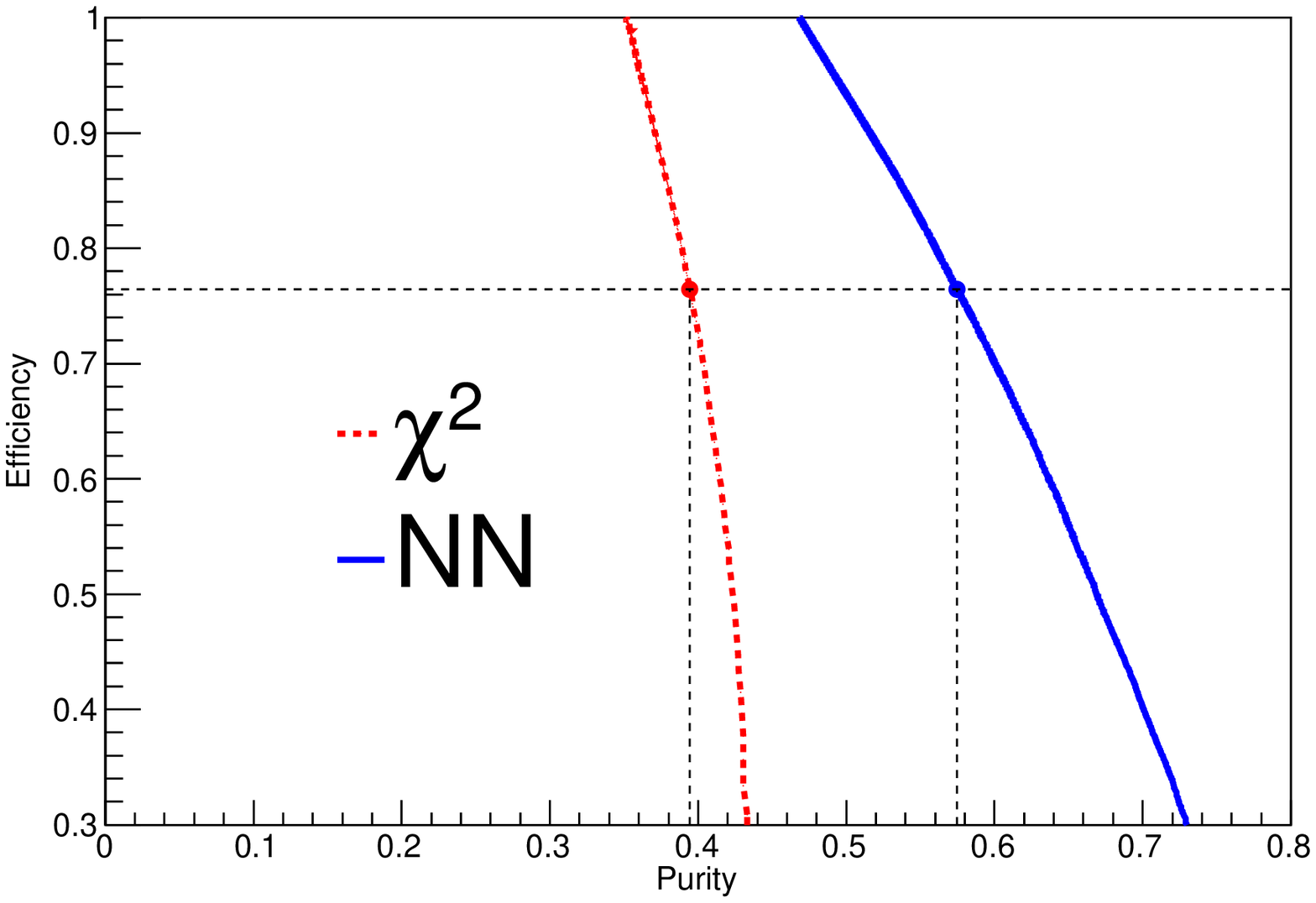}\\
(a) & (b) \\
\end{tabular}
\caption[Data fit]{(a) Distribution of the neural network output for the \afb measurement of the 
signal and the background using a SM $\ttbar$ sample with $\mtop=\gevcc{173}$. (b) The relationship 
between the purity and the efficiency of the NN method 
and the $\chi^2$ method is shown. At the same efficiency (76\%), which corresponds to the 
$\chi^2 <9$ cut, the NN method has 57\% purity, while the $\chi^2$ method has 39\% purity. 
}
\label{NN_afb_out}
\end{cfigure1c}

We show the purity as a function of the efficiency in Fig.~\ref{NN_afb_out} (b) for both 
the $\chi^2$ method and the NN method. We obtain the $\chi^2 <9$ efficiency when 
NN$_{\text{out}}>0.58$. In this condition, we obtain 57\% purity with the NN method, which
is approximately 46\% better than the $\chi^2$ method.

\begin{cfigure1c}
\begin{tabular}{cc}
\includegraphics[width=0.45\textwidth]{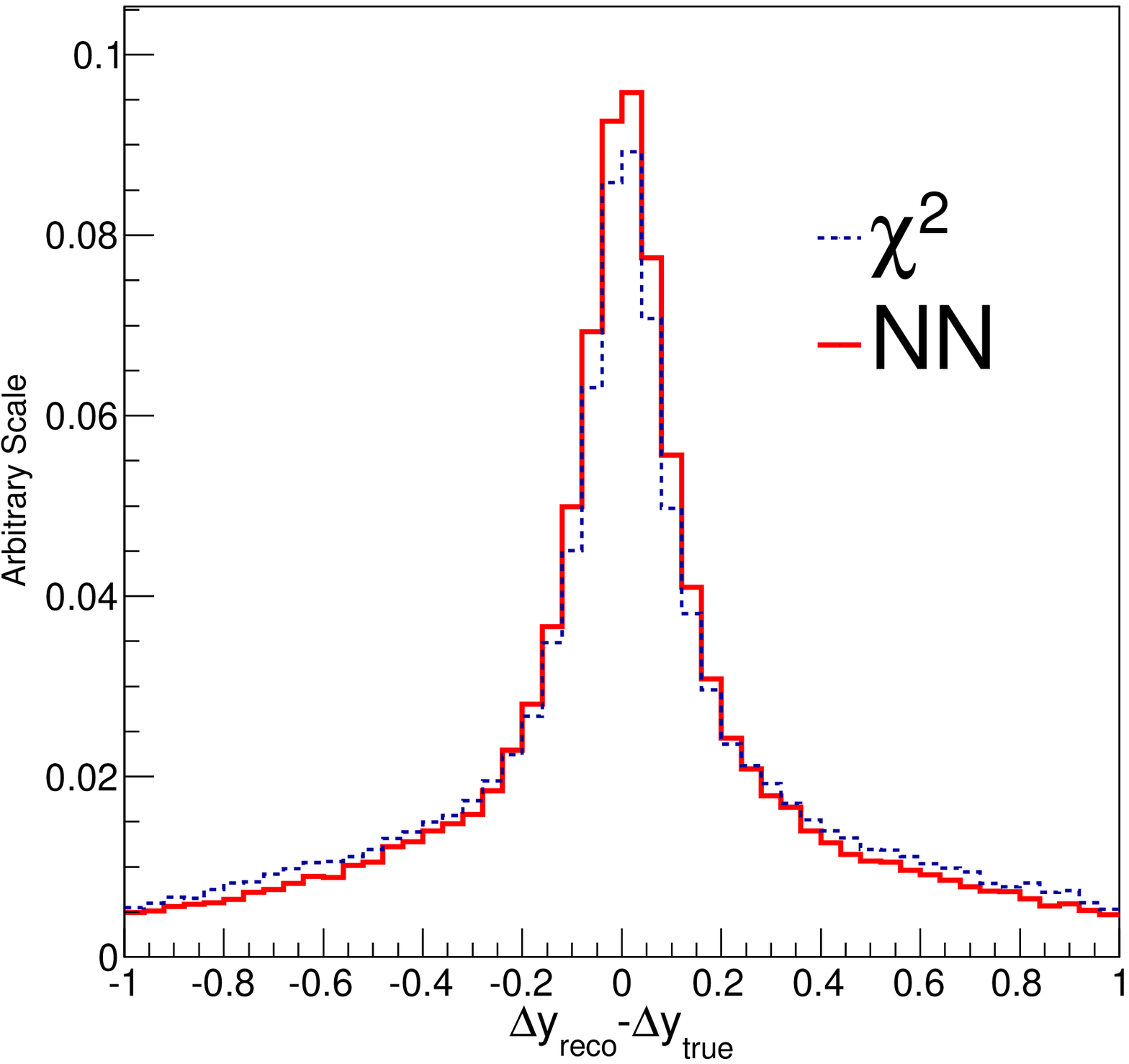}&
\includegraphics[width=0.45\textwidth]{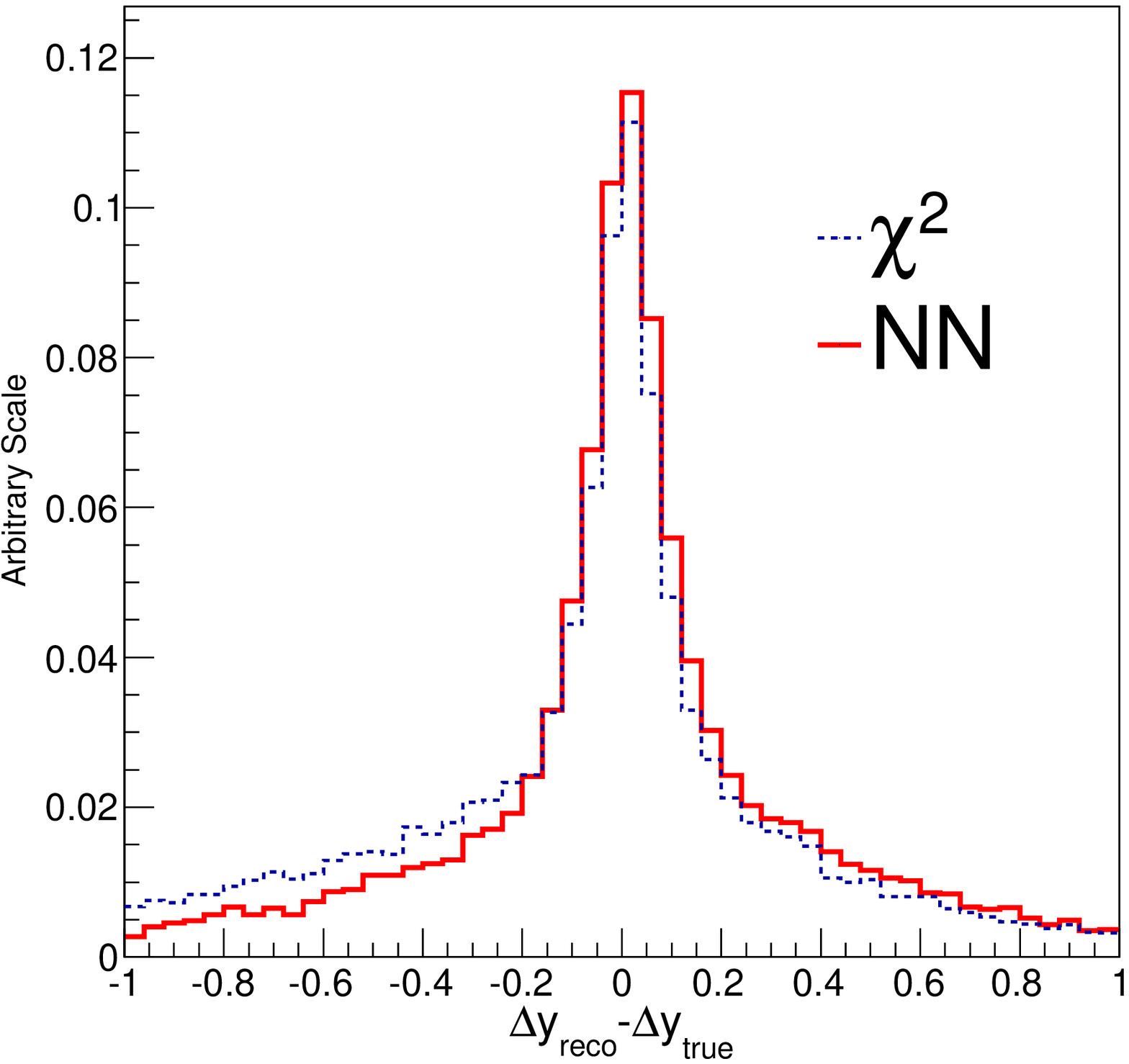}\\
(a) & (b) \\
\end{tabular}
\caption[Data fit]{ The $\Delta y_{\text{reco}}$ minus $\Delta y_{\text{true}}$ distributions 
of the (a) SM \ttbar sample and the (b) axigluon sample are shown using the NN method~(solid line) 
and the $\chi^2$ method~(dashed line). In both models, NN method has better resolution. 
}
\label{NN_angle_comp}
\end{cfigure1c}

In the measurement of $\afb$, the reconstructed rapidity difference between $t$ and 
$\bar{t}$~($\Delta y = y_{t}-y_{\bar{t}}$) are widely 
used~\cite{cdf_afb,cdf_afbfull,d0_afb,atlas_afb}. We investigate the reconstructed 
$\Delta y$~($\Delta y_{\text{reco}}$) using the NN method as well as the $\chi^2$ method to 
verify the effectiveness of the reconstruction method for real measurements. 
Figure~\ref{NN_angle_comp}~(a) shows $\Delta y_{\text{reco}}$ minus the true 
$\Delta y$~($\Delta y_{\text{true}}$) of the SM \ttbar sample. As we can see, the NN method 
has better resolution, which is approximately 9\% better than that of the $\chi^2$ method.

Because \afb is approximately zero in the \ttbar production of the LO SM process, 
we generate new physical processes with any significant $\afb$ value. Based on interesting 
models used to explain Tevatron \afb results, we use the axigluon~\cite{axigluon}~(with \tevcc{3}
mass) mediated top quark production. To generate the axigluon model, we use the 
{\sc madgraph/madevent} package with the top-BSM model~\cite{topBSM}. 
We apply the NN trained by the SM \ttbar sample and examine the $\Delta y$ distributions. 
As shown in Fig.~\ref{NN_angle_comp}~(b), $\Delta y_{\text{reco}}-\Delta y_{\text{true}}$ does
not shift with the NN method. In this sample, the true $\afb$ is $0.57$. We can also see that
the NN method has the better resolution, approximately 11\%, for the $\Delta y_{\text{reco}}$ 
distribution. 
Therefore, we can clearly improve \afb measurements at Tevatron using the NN method instead of 
the $\chi^2$ method. 

\section{Summary and conclusion}
In this study, we investigate the feasibility of using an artificial NN to resolve 
combinatorial issues in the \ttbar events at the hadron collider. We concentrated on the
lepton+jets decay topology where the four reconstructed jets should be matched with the 
four initial quarks. By including several input variables in the NN training, we have obtained 
very good discrimination between the signal and the background from NN$_\text{out}$. We then 
developed a reconstruction method based on NN$_\text{out}$. We have compared this method with 
the $\chi^2$ method and improved the purity by 21\% and 46\% for the \mtop and \afb measurements, 
respectively, without compromising the efficiency. We also present the reconstructed \mtop 
and $\Delta y$ distributions for the \mtop and \afb measurements, respectively. The NN does 
not introduce any additional bias compared with that of the $\chi^2$ method, but the resolutions 
of the reconstructed variables are significantly improved. We therefore conclude that the 
NN method can improve the precision of important top-quark measurements such as \mtop and \afb. 
We plan to revisit this method using a full detector simulation with experimental groups. 
 
The technique discussed in this paper is highly model dependent. However, multivariable 
techniques for performing event reconstruction can be applied to both well-known SM process 
measurements and to Beyond Standard Model (BSM) process measurements, if we have well-developed 
benchmark models. From this point of view, the technique discussed in this study can be a 
very powerful tool for resolving combinatoric ambiguities at the hadron collider. 

\begin{acknowledgments}
This research was supported by the Basic Science Research Program through the National Research 
Foundation of Korea (NRF) funded by the Ministry of Education~(NRF-2011-35B-C00007).
\end{acknowledgments}


\begin{thebibliography}{99}
	\bibitem{top_discovery} 
		F.~Abe {\it et al.} (CDF Collaboration), \emph{Observation of Top Quark Production in \ppbar  Collisions with the Collider Detector at Fermilab}, \emph{Phys. Rev. Lett.} {\bf 74} (1995) 2626; 
		S.~Abachi {\it et al.} (D0 Collaboration), \emph{Observation of the Top Quark}, \emph{Phys. Rev. Lett.}  {\bf 74} (1995) 2632.
	\bibitem{pdg} 
		J.~Beringer {\it et al.} (Particle Data Group), \emph{Review of Particle Physics}, \emph{Phys. Rev. D} {\bf 86} (2012) 010001.
	\bibitem{sbhigg} 
		M. Perelstein, M. E. Peskin, and A. Pierce, \emph{Top quarks and electroweak symmetry breaking in little Higgs models}, \emph{Phys. Rev. D} {\bf 69} (2004) 075002.
	\bibitem{sbreview} 
		G.~Bhattacharyya, \emph{Rept. Prog. Phys.}, \emph{A Pedagogical Review of Electroweak Symmetry Breaking Scenarios}, {\bf 74} (2011) 026201.
	\bibitem{tsearch}  
		T.~Aaltonen {\it et al.} (CDF Collaboration), \emph{Search for Resonant Top-Antitop Production in the Lepton Plus Jets Decay Mode Using the Full CDF Data Set}, \emph{Phys. Rev. Lett.} {\bf 110} (2013) 121802; 
		V. M. Abazov {\it et al.} (D0 Collaboration), \emph{Search for a Narrow \ttbar Resonance in \ppbar Collisions at $\sqrt{s}$=1.96 TeV}, \emph{Phys. Rev. D} {\bf 85} (2012) 051101; 
		G.~Aad {\it et al.} (ATLAS Collaboration), \emph{Search for \ttbar resonances in the lepton plus jets final state with ATLAS using \invfb{4.7} of $pp$ collisions at $\sqrt{s}$ = 7 TeV}, \emph{Phys. Rev. D} {\bf 88} (2013) 012004; 
		S.~Chatrchyan {\it et al.} (CMS Collaboration), \emph{Searches for new physics using the \ttbar invariant mass distribution in $pp$ collisions at $\sqrt{s}$ = 8 TeV}, \emph{Phys. Rev. Lett.} {\bf 111} (2013) 211804. 
	\bibitem{cdf_afb} 
		T.~Aaltonen {\it et al.} (CDF Collaboration), \emph{Evidence for a mass dependent forward-backward asymmetry in top quark pair production}, \emph{Phys. Rev. D} {\bf 83} (2011) 112003.
	\bibitem{d0_afb} 
		V.~M.~Abazov {\it et al.} (D0 Collaboration), \emph{Forward-backward asymmetry in top quark-antiquark production}, \emph{Phys. Rev. D} {\bf 84} (2011) 112005.
	\bibitem{cdf_afbfull} 
		T.~Aaltonen {\it et al.} (CDF Collaboration), \emph{Measurement of the top quark forward-backward production asymmetry and its dependence on event kinematic properties}, \emph{Phys. Rev. D} {\bf 87} (2013) 092002.
    \bibitem{elfit} 
		ALEPH, CDF, D0, DELPHI, L3, OPAL, SLD, LEP Electroweak Working Group, Tevatron 
		Electroweak Working Group, and SLD Electroweak and Heavy Flavor Working Groups, \emph{Precision Electroweak Measurements and Constraints on the Standard Mode}, arXiv:1012.2367.
    \bibitem{gfit} 
		H.~Fl$\ddot{\text{a}}$cher {\it et al.}, \emph{Revisiting the global electroweak fit of the Standard Model and beyond with Gfitter}, \emph{Eur.~Phys.~J.~C} {\bf 60} (2009) 543.
    \bibitem{higgs} 
		G.~Aad {\it et al.} (ATLAS Collaboration), \emph{bservation of a new particle in the search for the Standard Model Higgs boson with the ATLAS detector at the LHC}, \emph{Phys. Lett. B} {\bf 716} (2012) 1; 
		S.~Chatrchyan {\it et al.} (CMS Collaboration), \emph{Observation of a new boson at a mass of 125 GeV with the CMS experiment at the LHC}, \emph{Phys. Lett. B} {\bf 716} (2012) 30.
    \bibitem{tevave} 
		T. Aaltonen {\it et al.} (CDF and D0 Collaborations), \emph{Combination of CDF and DO results on the mass of the top quark using up to \invfb{8.7} at the Tevatron}, arXiv:1305.3929.
    \bibitem{lhcave} 
		CMS and ATLAS Collaborations, \emph{Combination of ATLAS and CMS results on the mass of the top quark using up to \invfb{4.9} of data}, CMS-PAS-TOP-13-005 and ATLAS-CONF-2013-102. 
    \bibitem{cdf_oldtop} 
		T. Aaltonen {\it et al.} (CDF Collaboration), \emph{Top quark mass measurement using the template method in the lepton+jets channel at CDF II}, \emph{Phys. Rev. D} {\bf 73} (2006) 032003. 
	\bibitem{d0_fitter} 
		B.~Abbott {\it et al.} (D0 Collaboration), \emph{Direct measurement of the top quark mass by the D0 Collaboration}, \emph{Phys. Rev. D} {\bf 58} (1998) 052001.
	\bibitem{atlas_mass} 
		G.~Aad {\it et al.} (ATLAS Collaboration), \emph{Measurement of the top quark mass with the template method in the \ttbar$\rightarrow$lepton+jets channel using ATLAS data}, \emph{Eur. Phys. J. C} {\bf 72} (2012) 2046.
	\bibitem{cms_mass} 
		S.~Chatrchyan {\it et al.} (CMS Collaboration), \emph{Measurement of the top-quark mass in \ttbar events with lepton+jets final states in $pp$ collisions at $\sqrt{s}$ = 7 TeV}, \emph{JHEP} {\bf 12} (2012) 105.
    \bibitem{comb1} 
		J. Alwall, K. Hiramastsu, M. M. Nojiri, and Y. Shimizu, \emph{Novel Reconstruction Technique for New Physics Processes with Initial State Radiation}, \emph{Phys. Rev. Lett.} {\bf 103} (2009)
			151802. 
    \bibitem{comb2} 
		A. Rajaraman and F. Yu, \emph{A new method for resolving combinatorial ambiguities at hadron colliders}, \emph{Phys. Lett. B} {\bf 700} (2011) 126. 
    \bibitem{comb3} 
		P. Baringer, K. Kong, M. McCaskey, and D. Noonan, \emph{Revisiting Combinatorial Ambiguities at Hadron Colliders with $M_{T2}$}, \emph{JHEP} {\bf 10} (2011) 101.
    \bibitem{NNhep1} 
		B.~Denby, \emph{Neural networks in high energy physics: A ten year perspective}, \emph{Comput. Phys. Commun.} {\bf 119} (1999) 219. 
    \bibitem{NNhep2} 
		L.~Teodorescu, \emph{Artificial neural networks in high-energy physics}, \emph{Proceedings of Inverted CERN School of Computing} 13 (2008).
	\bibitem{madgraph} 
		J.~Alwall {\it et al.}, \emph{MadGraph/MadEvent v4: the new web generation}, \emph{JHEP} {\bf 09} (2007) 028.
	\bibitem{pythia} 
		T.~Sjostrand {\it et al.}, \emph{High-energy-physics event generation with PYTHIA 6.1}, \emph{Comput. Phys. Commun.} {\bf 135} (2001) 238.
	\bibitem{pgs} 
		J.~Conway {\it et al.}, \emph{http://www.physics.ucdavis.edu/$\sim$conway/research/software/pgs/pgs4-general.htm}.
	\bibitem{cdf_fitter} 
		T.~Aaltonen {\it et al.} (CDF Collaboration), \emph{First simultaneous measurement of the top quark mass in the lepton+jets and dilepton channels at CDF}, \emph{Phys. Rev. D} {\bf 79} (2009) 092005; 
		T.~Aaltonen {\it et al.} (CDF Collaboration), \emph{Precision Top-Quark Mass Measurement at CDF}, \emph{Phys. Rev. Lett.} {\bf 109} (2012) 152003. 
	\bibitem{wmass} 
		T. Aaltonen {\it et al.} (CDF Collaboration), \emph{Precise Measurement of the $W$-Boson Mass with the CDF II Detector}, \emph{Phys. Rev. Lett.} {\bf 108} (2012) 151803; 
		V.~M.~Abzov {\it et al.} (D0 Collaboration), \emph{Measurement of the $W$ Boson Mass with the D0 Detector}, \emph{Phys. Rev. Lett.} {\bf 108} (2012) 151804.
	\bibitem{root} 
		R. Brun and F. Rademakers, \emph{ROOT-An object oriented data analysis framework,} \emph{Nucl. Instrum. Methods Phys. Res., Sect. A} {\bf 389} (1997) 81;
		See also \emph{http://root.cern.ch}.
	\bibitem{azidecorr} 
		S.~Choi and H.~S. Lee, \emph{Azimuthal decorrelation in \ttbar production at hadron colliders}, \emph{Phys. Rev. D} {\bf 87} (2013) 034012.
	\bibitem{atlas_afb} 
		G.~Aad {\it et al.} (ATLAS Collaboration), \emph{Measurement of the top quark pair production charge asymmetry in proton-proton collisions at $\sqrt{s}$ = 7 TeV using the ATLAS detector}, arXiv:1311.6724.
	\bibitem{axigluon} 
		P.~H.~Frampton and S.~L.~Glashow, \emph{Chiral color: An alternative to the standard model}, \emph{Phys. Lett. B} {\bf 190} (1987) 157; 
		J.~Bagger, C.~Schmidt, and S.~King, \emph{Axigluon production in hadronic collisions}, \emph{Phys. Rev. D} {\bf 37} (1988) 1188.
	\bibitem{topBSM} 
		R.~Frederix and F.~Maltoni, \emph{Top pair invariant mass distribution: a window on new physics}, \emph{JHEP} {\bf 01} (2009) 047.
\end{thebibliography}
\end{document}